# Pyramid diffractive optical networks for unidirectional image magnification and demagnification

Bijie Bai [1,2,3,†], Xilin Yang[1,2,3,†], Tianyi Gan[1,3], Jingxi Li[1,2,3], Deniz Mengu[1,2,3], Mona Jarrahi[1,3], and Aydogan Ozcan[*,1,2,3]

[1]Electrical and Computer Engineering Department, University of California, Los Angeles, CA, 90095, USA.

[2]Bioengineering Department, University of California, Los Angeles, 90095, USA.

[3]California NanoSystems Institute (CNSI), University of California, Los Angeles, CA, USA.

[†]Equal contributing authors

[*]Correspondence: Aydogan Ozcan. Email: ozcan@ucla.edu

# Abstract

Diffractive deep neural networks ($D^2$NNs) are composed of successive transmissive layers optimized using supervised deep learning to all-optically implement various computational tasks between an input and output field-of-view (FOV). Here, we present a pyramid-structured diffractive optical network design (which we term P-$D^2$NN), optimized specifically for unidirectional image magnification and demagnification. In this design, the diffractive layers are pyramidally scaled in alignment with the direction of the image magnification or demagnification. This P-$D^2$NN design creates high-fidelity magnified or demagnified images in only one direction, while inhibiting the image formation in the opposite direction – achieving the desired unidirectional imaging operation using a much smaller number of diffractive degrees of freedom within the optical processor volume. Furthermore, P-$D^2$NN design maintains its unidirectional image magnification/demagnification functionality across a large band of illumination wavelengths despite being trained with a single wavelength. We also designed a wavelength-multiplexed P-$D^2$NN, where a unidirectional magnifier and a unidirectional demagnifier operate *simultaneously* in opposite directions, at two distinct illumination wavelengths. Furthermore, we demonstrate that by cascading multiple unidirectional P-$D^2$NN modules, we can achieve higher magnification factors. The efficacy of the P-$D^2$NN architecture was also validated experimentally using terahertz illumination, successfully matching our numerical simulations. P-$D^2$NN offers a physics-inspired strategy for designing task-specific visual processors.

## Introduction

The fusion of machine learning techniques and optics/photonics has fostered major advancements in recent years, bridging the gap between traditional computational methods and the promising avenues of optical processing[1–3]. With the recent advances in data-driven design methodologies, optical computing platforms have gained design complexity with new capabilities, providing transformative solutions for various computational tasks[4–9]. These optical computing and visual processing platforms utilize the unique characteristics of light, such as phase, spectrum and polarization, to rapidly process optical information, offering advantages of parallel processing, computational speed, and energy efficiency. In this line of research, diffractive deep neural networks (D²NNs) have emerged as a free-space optical platform that leverages supervised deep learning algorithms to design diffractive surfaces for visual processing and all-optical computational tasks[10,11]. After their fabrication, these diffractive optical networks form physical processors of visual information, capable of executing various computer vision tasks, spanning image classification[10,12–15], quantitative phase imaging (QPI)[16,17], universal linear transformations[18–21], image encryption[22–24], and imaging through diffusive media[25,26], among many others[27–34]. The visual processing and optical computing capabilities of D²NNs hinge on the modulation of light diffraction through a sequence of spatially structured and optimized diffractive surfaces. Within the modulation area of each diffractive layer, there exist hundreds of thousands of light modulation units, each with a lateral feature size of ~$\lambda$/2, forming the diffractive neurons/features of the optical network, which represent the independent degrees of freedom of the visual processor. Complex-valued transmission coefficients of these diffractive layers are optimized using deep learning algorithms, and once fabricated, a D²NN completes its computational task at the speed of light propagation through passive light diffraction within a thin volume, making it a powerful tool for optical processing of visual information.

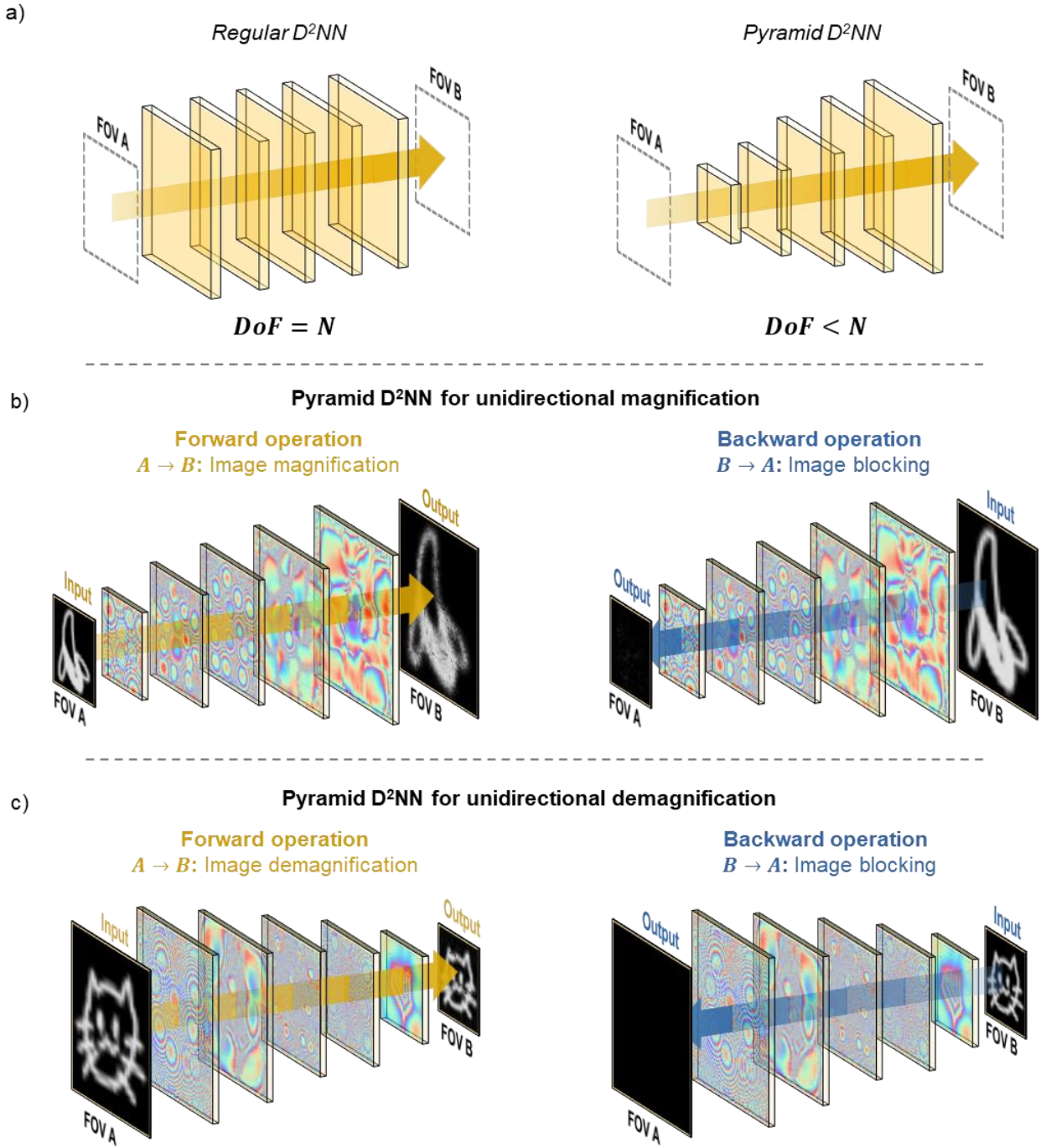

**Figure 1. Schematic of pyramid diffractive optical networks for unidirectional image magnification and demagnification.** (a) Comparison of a regular D²NN design and a P- D²NN design, where the P-D²NN has smaller degrees-of-freedom (DoF, i.e., the number of independent diffraction-limited features) than the regular D²NN. (b) P-D²NN for unidirectional image magnification. The diffractive network performs image magnification in the forward direction ($A \rightarrow B$) and image blocking in the opposite

direction ($B \rightarrow A$). (c) P-D$^2$NN for unidirectional image demagnification. The network performs image demagnification in the forward direction ($A \rightarrow B$) and image blocking in the opposite direction ($B \rightarrow A$).

---

Here, we present a pyramid-structured diffractive optical network design (Fig. 1a) and demonstrate its utility for unidirectional image magnification and demagnification tasks. In this pyramid diffractive network design (termed P-D$^2$NN), the size of the successive diffractive layers, and consequently, the number of diffractive neurons/features on each layer, scale in alignment with the desired magnification or demagnification factor. Therefore, the size of the initial diffractive layer is proportional to the size of the input object field-of-view (FOV), while the size of the terminal diffractive layer aligns with the size of the output FOV – following an image magnification or demagnification operation. Intermediate diffractive layers are proportionally scaled to geometrically align with the evolving fields during light propagation within the diffractive network volume (Figs. 1b-c). Based on this geometrical optics-inspired P-D$^2$NN architecture, we demonstrated unidirectional image magnification and demagnification tasks; when the incident light propagates along one pre-determined direction, the diffractive network magnifies (or demagnifies) the input images and generates the magnified (or demagnified) images at the output FOV. On the other hand, when the incident light propagates along the opposite direction, the diffractive network inhibits image formation, generating very low-intensity and unrecognizable images at the output FOV (Figs. 1b-c). We evaluated the effectiveness of the P-D$^2$NN architecture by comparing it against conventional D$^2$NN designs with uniform-sized diffractive layers. Our results indicate that P-D$^2$NN designs can achieve improved forward energy efficiency and stronger backward energy suppression for unidirectional image magnification/demagnification tasks compared to the performance of regular D$^2$NN architectures – using only half of the diffractive features due to their tapered geometry. Furthermore, our P-D$^2$NN-based unidirectional image magnifier/demagnifier designs maintain their functionality under a broad range of illumination wavelengths, even though they were trained using a single wavelength. We also designed a wavelength-multiplexed P-D$^2$NN that simultaneously performs *unidirectional magnification* at one wavelength of operation, while performing *unidirectional demagnification* at another wavelength in the opposite direction, further demonstrating the design versatility of the presented system.

Moreover, we demonstrated the cascadability of unidirectional P-D$^2$NNs, allowing for higher magnification factors by cascading multiple diffractive networks, each optimized to perform unidirectional image magnification. This modular approach is demonstrated by cascading two smaller unidirectional P-D$^2$NNs to achieve an enhanced overall magnification factor of $M = 3 \times 3 = 9$. This capability to cascade unidirectional P-D$^2$NNs demonstrates design flexibility to achieve various desired magnification factors by assembling multiple smaller diffractive modules.

We experimentally verified the efficacy of our P-D²NN framework using monochromatic terahertz (THz) illumination. After its deep learning-based optimization, the resulting diffractive layers were fabricated using 3D printing and assembled to be tested under continuous-wave THz illumination at λ = 0.75 mm. We experimentally validated the efficacy of the unidirectional P-D²NN framework using three different designs: two unidirectional magnifier designs with magnification factors of $M = 2$ and $M = 3$, and a unidirectional demagnifier with a demagnification factor of $D = 2$. All the experimentally measured results closely match our numerical simulations, where the output images in the forward direction accurately reflect the magnified or demagnified versions of the input images, while the outputs in the opposite (backward) direction produce low-intensity, non-informative results – as desired from a unidirectional imager.

As a unidirectional imaging system capable of magnifying or demagnifying images, the P-D²NN framework not only suppresses backward energy transmission but also disperses the original signal into unperceivable noise at the output of the backward direction. This unidirectional imaging capability cannot be achieved using standard lens designs, and, together with its polarization-insensitive operation, it could be of broad interest for various applications, including optical isolation for photonic devices, decoupling of transmitters and receivers for telecommunication, privacy-protected optical communications and surveillance. As another example of a potential application, P-D²NNs can be designed to deliver high-power structured beams onto target objects independent of the input polarization state, while protecting the source from counter-attacks or external beams. Compared to the standard, uniformly-sized D²NNs, this physics-inspired pyramid diffractive network architecture utilizes significantly fewer diffractive features per design, which is important to mitigate potential data overfitting issues and reduce fabrication costs in the deployment of visual processors, covering various applications in e.g., computer vision, robotics, and autonomous systems.

## Results

### P-D²NN for unidirectional image magnification and demagnification

Throughout this study, we refer to the optical path from FOV A to FOV B as the forward direction, and the reverse path as the backward direction (see Fig. 1a). We first demonstrate unidirectional image magnification using a spatially coherent pyramid diffractive optical network, as illustrated in Fig. 1b. In this optical system, when the incident light propagates along the forward direction, the diffractive network magnifies the input images from FOV A and generates the corresponding magnified output images at FOV B. However, as a *unidirectional* image magnification system, the opposite path functions differently. When images at FOV B propagate along the backward direction, the diffractive network inhibits the image formation at FOV A by scattering the optical fields outside of the output FOV, therefore resulting in very low-intensity and unrecognizable output images at FOV A – as desired in a unidirectional imaging design.

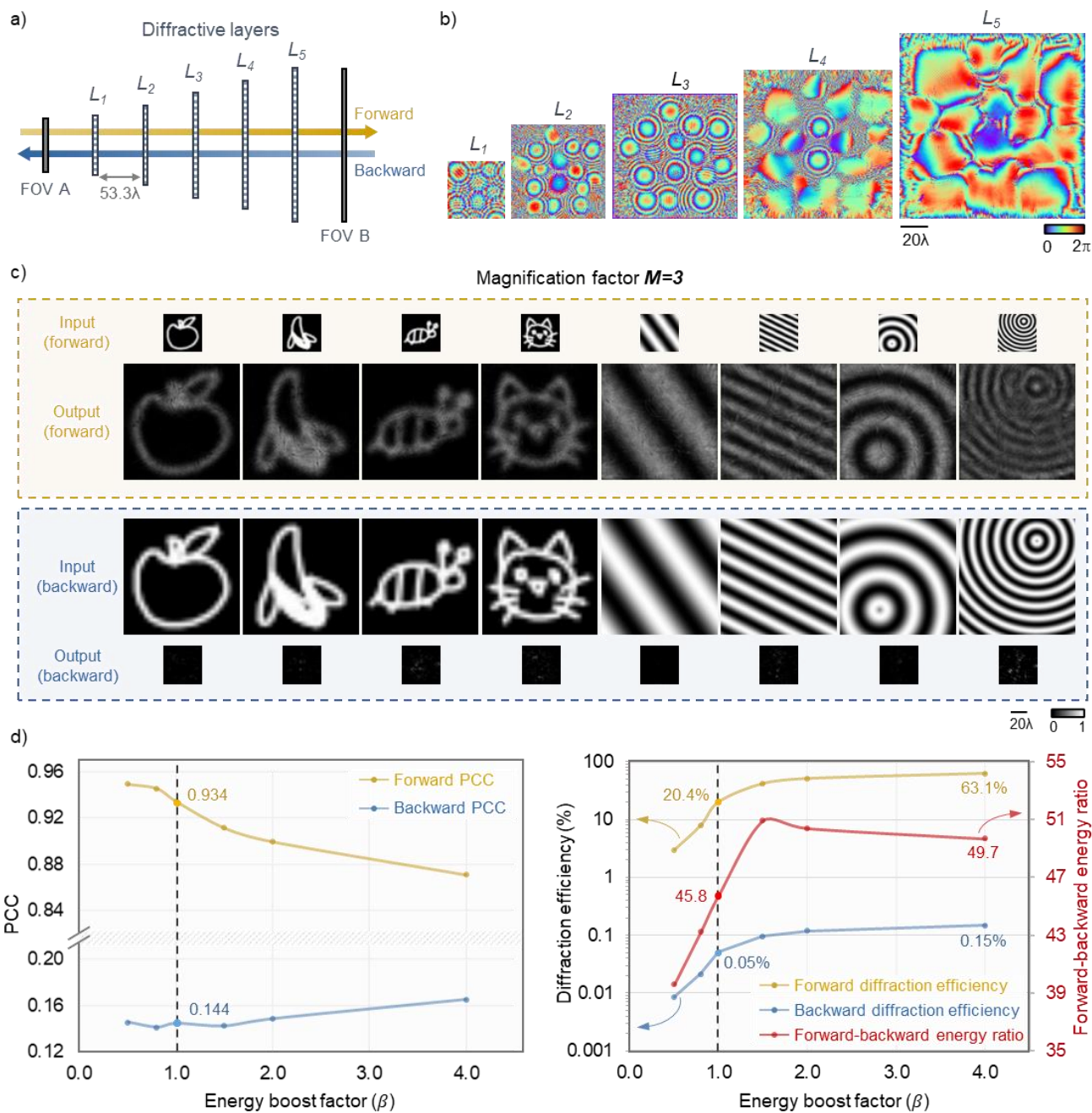

**Figure 2. Design schematic and blind testing results of the pyramid unidirectional image magnifier.** (a) Layout of a five-layer P-D²NN for unidirectional image magnification. (b) The resulting diffractive layers after deep learning-based optimization (with an energy boost factor of $\beta = 1$). (c) Examples of blind testing results of the trained unidirectional image magnifier in both the forward and backward

directions ($\beta = 1$). (d) Quantitative evaluations of six independent unidirectional image magnifiers trained with varying $\beta$ values. Each data point is the average from 1600 test images.

As shown in Fig. 2a, the pyramid network used for unidirectional image magnification contains five diffractive layers with progressively increasing numbers of diffractive features on each layer. These diffractive features on each surface have a characteristic size of approximately half the wavelength of the illumination light, which modulate the phase of the transmitted optical field by introducing an optical path length difference at the diffraction limit of light. Outside the effective areas of the diffractive layers that contain these phase modulation features, the regions at the edges are set as non-transmissive, completely blocking the light field that reaches these edge regions of a diffractive layer. This P-D$^2$NN architecture is designed to achieve a geometrical image magnification factor of $M = 3$ in the forward direction. In this configuration, the size of the progressively increasing diffractive layers are set to 90×90, 140×140, 180×180, 220×220, and 270×270 pixels (diffractive features), respectively, leading to a total number of $N = N_b = 181,400$ trainable diffractive neurons. The axial spacing between consecutive layers was set to ~53.3$\lambda$.

Based on this geometric configuration, the pyramid diffractive network was first digitally modeled, and the modulation depths of all diffractive features were iteratively optimized using deep learning (see the Methods section). The optimization target was driven by minimizing a set of custom-designed loss functions that enable unidirectional image magnification, designed to achieve three primary objectives: (1) maximizing the structural similarity between the output images in the forward direction ($A \rightarrow B$) and the corresponding ground truth images (i.e., the magnified versions of the input images) using normalized mean square error (NMSE) and the negative Pearson Correlation Coefficient (PCC)[35]; (2) enhancing the diffraction efficiency in the forward direction ($A \rightarrow B$); and (3) suppressing the diffraction efficiency in the backward direction ($B \rightarrow A$). Further details of the network architecture and the mathematical formulation of the loss functions can be found in the Methods section. Utilizing these customized loss functions, the optimization of the diffractive layers was carried out via a data-driven supervised training process using the images from the QuickDraw dataset[36] supplemented by an additional image dataset with grating/fringe-like patterns[17,29]. By tuning the weighting coefficient (i.e., energy boost factor $\beta$) of the customized loss term designed for enhancing the diffraction efficiency in the forward direction ($A \rightarrow B$), the diffractive networks were successfully trained to simultaneously achieve high-quality image magnification and a decent diffraction efficiency in the forward direction (see Figs. 2c-d). In our quantitative performance analyses, we trained six independent models with the same P-D$^2$NN architecture using different $\beta$ values (see the Methods section). These models were subsequently tested on a separate dataset of 1600 test images which were not seen during the training phase. The performance of each trained P-D²NN was quantified based on several metrics: (1) PCC between the output images and the corresponding ground truth images (i.e., the magnified input images) in the forward direction ($A \rightarrow B$); (2) PCC between the output images and the corresponding ground truth images (i.e., the demagnified input images) in the backward direction ($B \rightarrow A$); (3) diffraction efficiency in the forward direction; (4) diffraction efficiency in the backward direction; and (5) the energy ratio between the forward and backward output images (see the Methods section). For example, Fig. 2b illustrates the diffractive layers of a converged P-D²NN trained using $\beta = 1$, whose blind test results are demonstrated in Fig. 2c. The quantitative metrics listed above were calculated for all $\beta$ settings, as summarized in Fig. 2d. For the $\beta = 1$ case, it is observed that the trained P-D²NN exhibits an asymmetric behavior, as desired, where the output images at the forward direction closely resemble the magnified input images with a PCC value as high as 0.934, and a forward diffraction efficiency of 20.4% (dashed lines in Fig. 2d). In contrast, the backward path only retains a diffraction efficiency of 0.05%, resulting in very low-intensity images with a backward PCC as low as 0.144 (Fig. 2d). This diffractive network achieves an average energy suppression ratio of ~46-fold between the backward and the forward directions, demonstrating the success of its unidirectional magnification.

Additional quantitative assessments across all six models with different $\beta$ values (Fig. 2d) reveal that increasing $\beta$ further boosts the forward diffraction efficiency. However, this enhancement is coupled with a decrease in the forward PCC and a slight increase in the backward PCC. The diffraction efficiency in the backward direction also increases slowly with larger $\beta$ values. As shown in Fig. 2d, the forward-backward energy ratio is first improved and then slowly drops beyond $\beta = 1.5$. Nonetheless, diffractive models with high energy efficiency can be designed without a significant decrease in the unidirectional imaging performance. For example, diffraction efficiency can be improved up to 51.4% with $\beta = 2$ while the unidirectional image magnification performance remains at a very good level (PCC = 0.9). Visualization of the blind testing image examples for different $\beta$ values can be found in Supplementary Fig. S1. We further trained and tested the P-D²NN framework with varying numbers of diffractive layers (denoted by $K$) from $K = 2$ to $K = 5$ maintaining an energy boost factor of $\beta = 4$. The blind testing results, summarized in Supplementary Fig. S2, indicate that an increased number of diffractive layers, as expected, improves the unidirectional imaging performance of P-D²NNs; also see the Methods section. These quantitative analyses and comparisons reveal that various design choices can adjust the P-D²NN design to achieve a desirable range of forward diffraction efficiency and unidirectional image magnification quality, while also significantly suppressing the backward PCC and diffraction efficiency (see Fig. 2d).

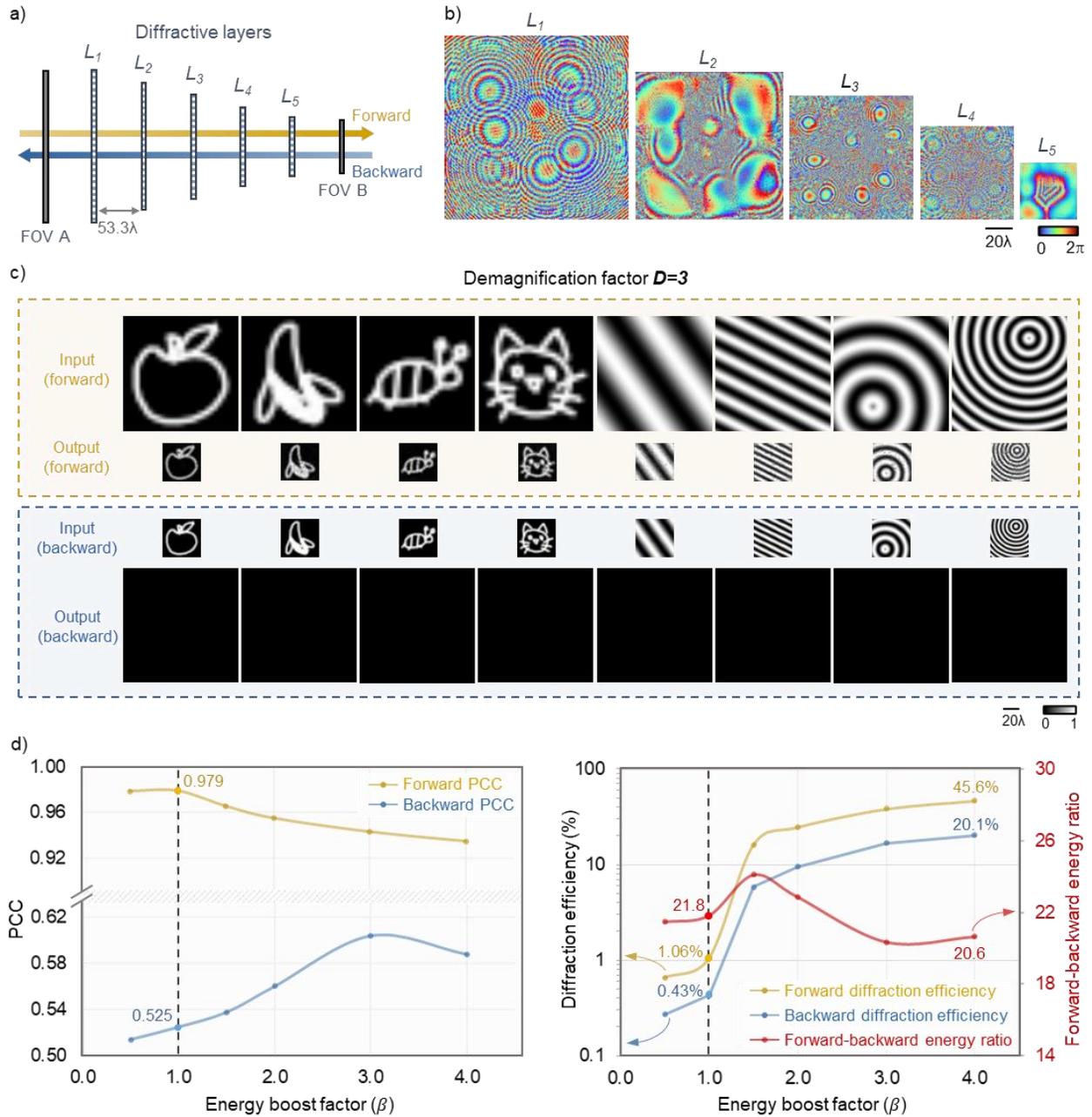

**Figure 3. Design schematic and blind testing results of the pyramid unidirectional image demagnifier.** (a) Layout of a five-layer P-D$^2$NN for unidirectional image demagnification. (b) The resulting diffractive layers after deep learning-based optimization ($\beta = 1$). (c) Examples of blind testing results of the trained unidirectional image demagnifier in both the forward and backward directions ($\beta = 1$). (d) Quantitative evaluations of six independent unidirectional image demagnifiers trained with varying $\beta$ values. Each data point is the average from 1600 test images.

To further investigate the imaging quality of the P-D$^2$NN framework, we conducted additional blind testing using various gratings and slanted edges (see the Methods). For this analysis, we tested a series of gratings with different periods, shifting them to 9 positions in a 3×3 grid within the input FOV, in both the forward and backward directions, to study the system's resolution and aberrations. The results are summarized in Supplementary Figs. S3 and S4, where our P-D$^2$NN design resolved gratings with a period of 4$\lambda$ and partially resolved gratings with a period of 3$\lambda$ – all in the forward direction. In the backward direction, the imaging is blocked, leaving no observable grating patterns – as desired. Additionally, a slanted-edge test was conducted with 9 rotation angles, both in the forward and backward directions, as summarized in Supplementary Fig. S5. The slanted edges are clearly imaged in the forward direction and blocked in the backward direction, demonstrating the unidirectional imaging capability of the P-D$^2$NN framework. To estimate the point-spread function (PSF) in each direction, we calculated the gradients of the image cross-sections perpendicular to the edges, which revealed a full-width at half maximum (FWHM) of 6.52$\lambda$ (see the Methods section). These results can be further improved by including objects with higher-resolution spatial features during the training process.

Following a similar design method, we also performed *unidirectional image demagnification* through a pyramid diffractive network with decreasing layer sizes along the forward light propagation direction, as illustrated in Fig. 1c. This diffractive network shrinks the input images at FOV A, yielding demagnified output images at FOV B along the forward path. Based on its unidirectional imaging design, the network inhibits the image formation from FOV B to FOV A in the backward direction and produces very weak and unrecognizable output images. Similar to the magnification P-D$^2$NN, this P-D$^2$NN design for unidirectional image demagnification comprises five diffractive layers, each containing progressively smaller number of diffractive features that modulate the phase of the transmitted optical fields (Fig. 3a). We selected a demagnification factor of $D$ = 3 in the forward direction. The axial spacing between successive diffractive layers is kept the same as before, ~53.3$\lambda$. The optimization process of the diffractive layers follows the same methodology as the unidirectional image magnification models reported in Fig. 2, where the same set of loss functions and training data were used (see the Methods section). The same quantitative analysis was also performed for the unidirectional image demagnification P-D$^2$NN using six unique models numerically trained under different energy boost factors $\beta$, and blindly tested using 1600 test images not included in the training phase, as summarized in Figs. 3b-d.

Figure 3b shows the diffractive layers of a converged P-D$^2$NN model designed with $\beta$ = 1, whose blind testing results are shown in Fig. 3c. The same asymmetric behavior is observed for the trained P-D$^2$NN, i.e., the output images in the forward direction are nearly identical to the demagnified versions of the input images, attaining a forward PCC of 0.979 and a forward diffraction efficiency of 1.06% (dashed lines in Fig. 3d), whereas the backward path only reaches a PCC of 0.525 and a diffraction efficiency of 0.43%, resulting in nearly dark output images. It is worth noting that the output images in both the forward and the backward directions, as depicted in Fig. 3c, are displayed with an identical range and the same color map. Although the forward and backward diffraction efficiencies, computed based on the total energy at their respective FOVs, might appear close, there is a substantial difference in the corresponding brightness of the forward/backward images due to the fact that the output images in the backward direction have much weaker average intensity per pixel (see Fig. 3c). In fact, Fig. 3d reveals that by varying the $\beta$ value used in the training, the forward diffraction efficiency values of this unidirectional demagnifier P-D²NN design can be increased to >45% with a very good forward PCC value of >0.92, while also suppressing the backward diffraction efficiency and backward PCC values to ≤ ~20% and < ~0.6, respectively. Visualization of the blind testing examples of the unidirectional demagnification P-D²NN designs trained with different $\beta$ values can be found in Supplementary Fig. S6.

**Comparison of P-D²NN performance against a regular D²NN architecture**

Next, we compare the performance of the P-D²NN architecture against a regular D²NN structure for unidirectional image magnification tasks. In this comparison, the P-D²NN model is directly taken from the unidirectional image magnification model trained with $\beta = 1$, as reported in Figs. 2b-c, which has a total of $N_b = 181,400$ diffractive features. The regular D²NN design employs uniform-sized diffractive layers, with the size of each layer equal to 270×270 pixels, yielding a total of $N = 2N_b$ trainable diffractive features. This standard D²NN design was trained using the same training loss functions (with $\beta = 1$), image datasets, and the number of epochs as we used for its pyramid counterpart shown in Figs. 2b-c. After its training, the blind inference was performed using the same test dataset of 1600 images to conduct the quantitative performance evaluations.

Figures 4a-b show the comparative blind testing results for the P-D²NN and the regular D²NN designs. In the forward direction, both diffractive networks demonstrate similar image magnification fidelity, as evident from both the visual assessments and the quantitative PCC values. This underscores the efficiency of the P-D²NN framework, which achieves similar performance levels using only about half as many diffractive features as the regular diffractive network design. Furthermore, the P-D²NN surpasses the standard diffractive model in terms of forward energy efficiency and energy suppression ratio, producing brighter images in the forward direction with significantly less energy in the backward direction, demonstrating a superior unidirectional imaging capability.

To further shed light on this comparison, we took the diffractive layers of the trained regular D²NN model and added light-blocking regions to each layer (Fig. 4c) with the sizes and the positions of the transmissive regions at each diffractive layer matching the corresponding layers in the P-D$^2$NN design. This "trimmed" D$^2$NN model, with only the central region of each diffractive layer participating in the inference process, has the same number of diffractive features as the P-D$^2$NN (i.e., $N = N_b$) and was benchmarked using the same 1600-image test dataset. Naturally, the performance of this trimmed D$^2$NN degrades compared to its original model, given that the peripheral diffractive neurons were disabled during the inference process. Moreover, when compared against the P-D$^2$NN model, this trimmed D$^2$NN produced inferior results across all image performance criteria (see Fig. 4c). This suggests that simply trimming an already-trained diffractive optical network to emulate the light propagation cone is not an effective approach.

To further investigate the influence of the diffractive layer dimensions on the performance of pyramid diffractive networks, we conducted additional analyses, where we adopted the P-D$^2$NN delineated in Figs. 2b-c as our baseline model (also shown in Supplementary Fig. S7a). From this baseline, we incrementally enlarged the dimensions of each diffractive layer by $m$ pixels, transforming, for instance, a $90 \times 90$ layer to $(90 + m) \times (90 + m)$, and a $270 \times 270$ layer to $(270 + m) \times (270 + m)$. For this analysis, we considered $m$ values of 20, 40, and 70 (as illustrated in Supplementary Figs. S7b-d). Consequently, P-D$^2$NN configurations with $N = 1.2N_b$, $N = 1.4N_b$, and $N = 1.8N_b$ were successively trained and quantitatively evaluated, with the results summarized in Supplementary Fig. S7. These analyses reveal that by infusing additional degrees of freedom into a P-D$^2$NN architecture, there is a modest improvement in the unidirectional imaging performance. Notably, in the case of $m = 70$ and $N = 1.8N_b$, P-D$^2$NN outperforms the regular D$^2$NN ($N = 2N_b$) in every quantitative performance metric, including higher PCC and diffraction efficiency in the forward direction. These findings further underscore the pyramid diffractive network configuration's architectural superiority for learning unidirectional image magnification (or demagnification) tasks.

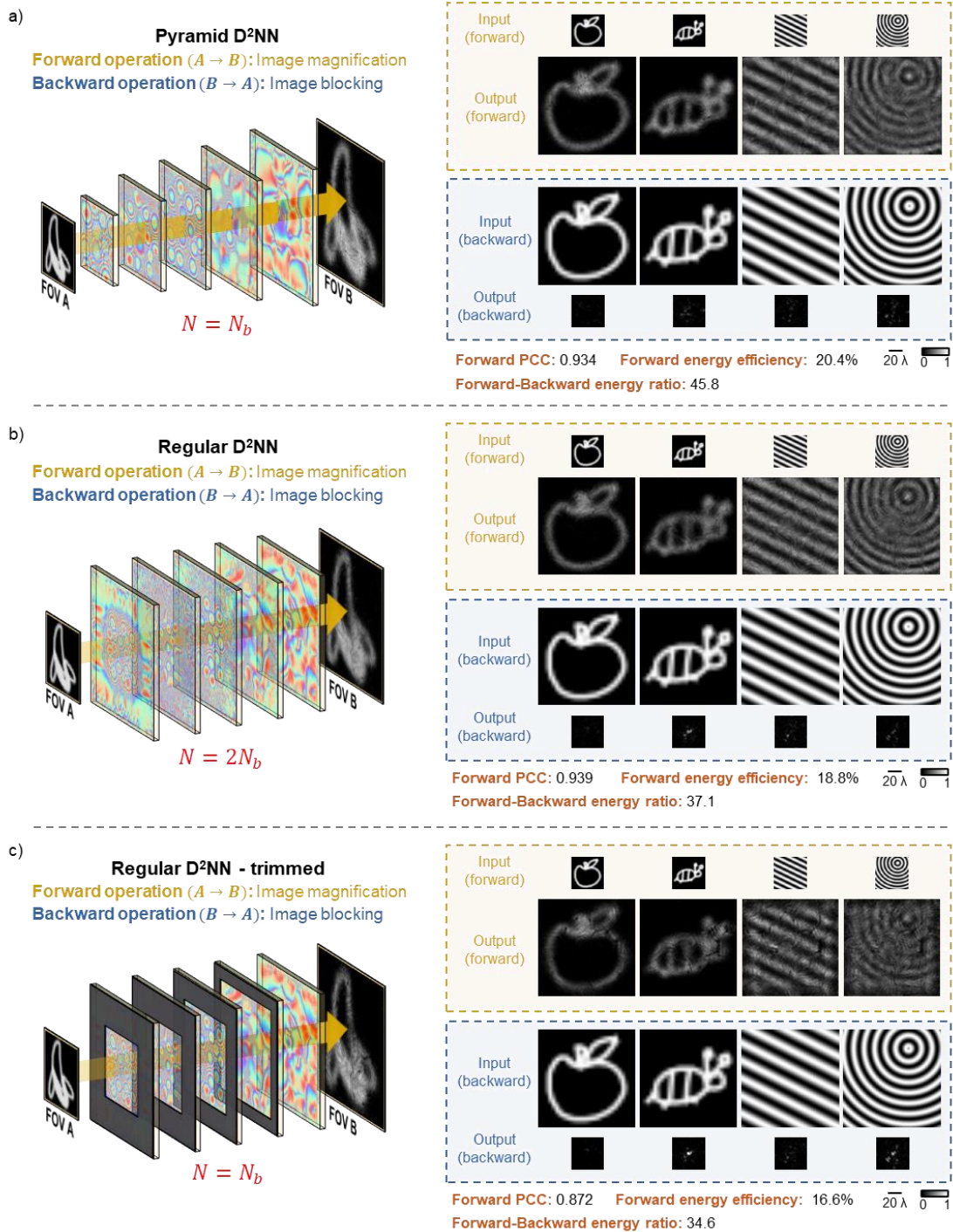

**Figure 4. Comparison of P-D²NN against a uniform-sized D²NN.** (a) Design layout and blind testing results of a P-D²NN-based unidirectional image magnifier, where the P-D²NN has $N = N_b$ independent diffractive features. (b) Design layout and blind testing results of a regular D²NN-based unidirectional magnifier, where the D²NN has $N = 2N_b$ independent diffractive features. (c) Design layout and blind testing results of a trimmed version of the regular D²NN, where the trimmed D²NN has $N = N_b$ diffractive features. The trimmed D²NN was obtained by taking the diffractive layers of the regular D²NN

(depicted in b) and adding light-blocking regions to match the transmissive regions of the P-D²NN (depicted in a).

---

**Spectral response of the pyramid unidirectional image magnification network**

Next, we investigated the spectral behavior of the pyramid unidirectional image magnifier depicted in Figs. 2b-c. This was done by taking the P-D²NN, initially trained at $\lambda_{\text{train}} = 0.75$ mm (Fig. 2b), and blindly testing it at a range of illumination wavelengths ($\lambda_{\text{test}}$) that diverged from the original training wavelength to assess its performance beyond the original training wavelength. Blind testing results for both the forward and backward paths across different $\lambda_{\text{test}}$ values are shown in Fig. 5a. Notably, although the unidirectional image magnifier P-D²NN was trained exclusively under a single illumination wavelength $\lambda_{\text{train}}$, it preserves its designed functionality over an extended spectral range, consistently achieving unidirectional image magnification in the forward path while suppressing image formation in the reverse path.

We further evaluated the generalization of the trained unidirectional image magnifier P-D²NN using a unique image dataset featuring resolution test targets with varying linewidths (Fig. 5b). The blind testing results at $\lambda_{\text{test}} = \lambda_{\text{train}}$ and $\lambda_{\text{test}} \neq \lambda_{\text{train}}$ validate the efficacy of P-D²NN in achieving a general-purpose unidirectional image magnifier, even though it was exclusively trained on a different dataset. These analyses demonstrated that the trained P-D²NN unidirectional magnifier can resolve a minimum linewidth of approximately $6.3\lambda$ when working in the forward direction ($A \rightarrow B$), while effectively suppressing image formation in the reverse direction, $B \rightarrow A$.

A comprehensive quantitative analysis is also presented in Fig. 5c, summarizing the blind testing performance metrics evaluated within an illumination band covering from $\lambda_{\text{test}} = 0.6$ mm to $\lambda_{\text{test}} = 0.9$ mm. These quantitative results reveal that, when operating in the forward path, the unidirectional magnifier maintains a high PCC value of $\geq 0.82$ within a spectral range of $[0.87\lambda_{\text{train}}, 1.17\lambda_{\text{train}}]$, i.e., within $[0.65$ mm, $0.88$ mm$]$. Its forward diffraction efficiency remains fairly stable ($\geq 17.8\%$) across the tested spectral range. In the reverse direction, on the other hand, the forward-backward energy ratio is maintained to be $\geq 20$ (and $\geq 30$) within a spectral range of $[0.89\lambda_{\text{train}}, 1.18\lambda_{\text{train}}]$ (and $[0.92\lambda_{\text{train}}, 1.11\lambda_{\text{train}}]$), respectively, demonstrating the broadband operation of this P-D²NN unidirectional magnifier design, although it was trained using a single illumination wavelength.

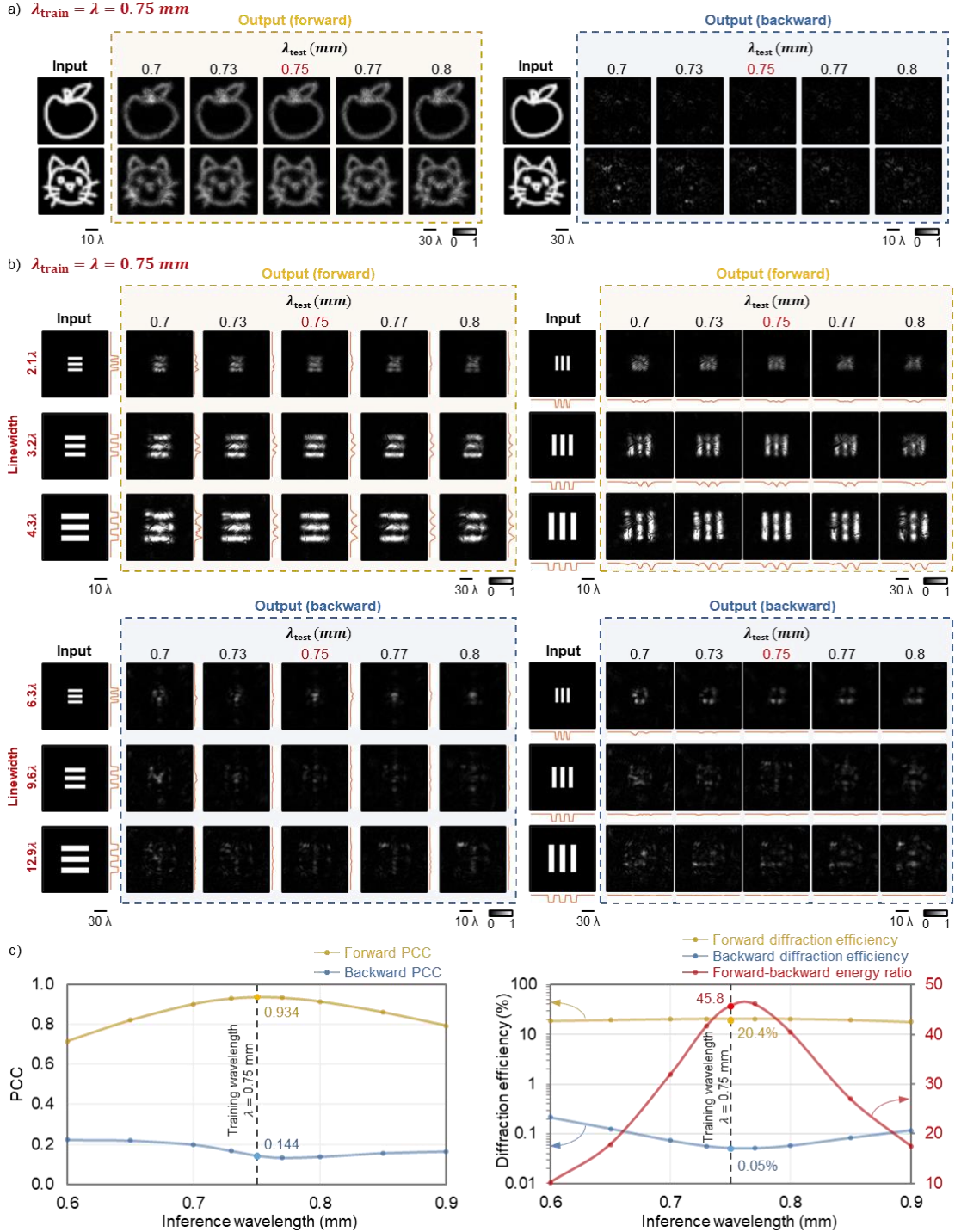

**Figure 5. Spectral response of the pyramid unidirectional image magnifier.** (a) Blind inference results when testing the unidirectional image magnifier model (trained using a single wavelength, as shown in Figs. 2b-c) at different illumination wavelengths. (b) Blind inference results when testing the same unidirectional magnifier model at different illumination wavelengths using customized resolution target images, demonstrating its generalization capability to new types of objects. (c) Quantitative

evaluation results of the same unidirectional image magnifier blindly tested at different illumination wavelengths. Each data point is the average from 1600 test images.

---

**Wavelength-multiplexed P-D²NN design for unidirectional image magnifier and demagnifier**

Next, we integrated the functions of a diffractive unidirectional magnifier and a diffractive unidirectional demagnifier into the same P-D²NN, but in the opposite directions. The directionality of magnified or demagnified imaging is determined by the illumination wavelength, as depicted in Fig. 6a. At an illumination wavelength of $\lambda_1$, the P-D²NN serves as a unidirectional *magnifier* in the forward direction, where the input images at FOV A are magnified at FOV B. Concurrently, the image formation is inhibited at $\lambda_1$ in the backward path from FOV B to FOV A. In contrast, at an illumination wavelength of $\lambda_2$, the image formation is inhibited in the forward path from FOV A to FOV B, while the image *demagnification* is achieved in the backward path, shrinking the images from FOV B to FOV A. For this wavelength-multiplexed design, we set $\lambda_1 = 0.75$ mm and $\lambda_1 = 0.80$ mm, and incorporated the same set of training loss functions as described before for $\lambda_1$ and $\lambda_2$ separately (with $\beta = 1$; see the Methods section). Upon completion of the training, the P-D²NN model underwent blind testing using a test set composed of 1600 unique images (see Fig. 6b for some examples). These visual evaluations demonstrate that the wavelength-multiplexed P-D²NN simultaneously performs two distinct unidirectional image scaling operations in opposite directions, with the directionality of the unidirectional imaging determined by the illumination wavelength. In the forward path, the image magnification function operates at $\lambda_1$, but remains inactive at $\lambda_2$. Conversely, in the backward path, the image demagnification function operates at $\lambda_2$ but remains inactive at $\lambda_1$ (see Fig. 6b).

We further trained and tested four wavelength-multiplexed unidirectional P-D²NN models with different energy boosting factors, i.e., $\beta = 2.5, 3, 4,$ and 5. The quantitative assessment of these different P-D²NN models is illustrated in Fig. 6c, showing the image *magnification* PCC and the diffraction efficiency in the forward direction $(A \rightarrow B)$ for $\lambda_1$, and the image *demagnification* PCC and the diffraction efficiency in the backward direction $(B \rightarrow A)$ for $\lambda_2$. These results indicate that the tuning of $\beta$ values during the training of these wavelength-multiplexed P-D²NNs can be used to adjust the trade-off between the image quality and the diffraction efficiency, simultaneously applicable for the magnification and demagnification functions at both operating wavelengths (see Fig. 6c).

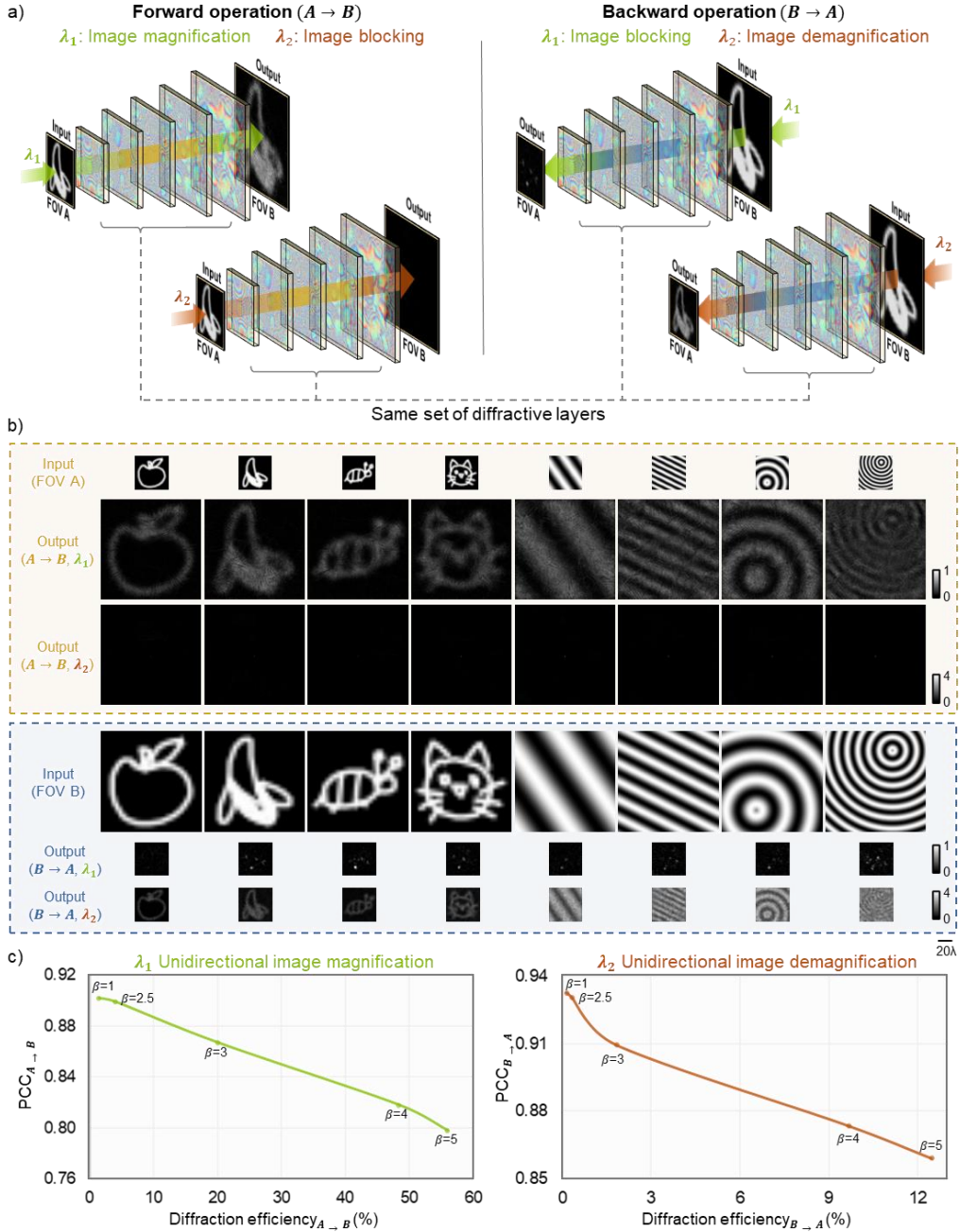

**Figure 6. Demonstration of a wavelength-multiplexed P-D²NN performing unidirectional image magnification and unidirectional image demagnification *simultaneously* at two distinct wavelengths.** (a) The design concept of the wavelength-multiplexed pyramid diffractive network. At $\lambda_1$, the network performs image magnification in its forward direction ($A \rightarrow B$) and image blocking in its backward direction ($B \rightarrow A$). Oppositely, at $\lambda_2$, the network performs image blocking in its forward direction ($A \rightarrow B$) and image demagnification in its backward direction ($B \rightarrow A$). (b) Examples of blind testing results of the wavelength-multiplexed P-D²NN in both the forward and backward directions at two distinct

wavelengths. (c) Quantitative comparison of various wavelength-multiplexed P-D²NN designs trained under different $\beta$ values, showing the trade-off between image magnification/demagnification fidelity and the corresponding diffraction efficiency along the same direction.

## Cascaded P-D²NNs to achieve higher magnification factors for unidirectional imaging

Next, we demonstrate that cascading unidirectional magnification diffractive networks can achieve a higher overall magnification through joint optimization. Figure 7a illustrates the structure of a cascaded P-D²NN where two smaller diffractive models achieve a cumulative magnification factor of $M = 3 \times 3 = 9$. This cascaded structure consists of two P-D²NNs, P1 and P2, each with four diffractive layers, where each subsequent layer is larger than the previous. The input and output apertures of P1 are defined as FOV A and B, respectively, with FOV B also serving as the input aperture for P2, whose output is denoted as FOV C. These three FOVs are color-coded and drawn to scale in Fig. 7b. Details of the structural parameters are provided in the Methods section.

To optimize the cascaded P-D²NN structure, we employed a joint optimization strategy. In this scheme, the total loss function is composed of three parts: the unidirectional magnification loss for each individual component (P1 and P2), and a third unidirectional magnification loss for the end-to-end optimization of the entire cascaded unit, as detailed in the Methods section. We conducted joint testing of the entire cascaded network, targeting an overall magnification factor of $M = 3 \times 3 = 9$, to evaluate its unidirectional imaging capabilities. The results, depicted in Fig. 7c, reveal that in the forward direction, the cascaded P-D²NN network created output images that closely align with the magnified input image – as desired. Conversely, in the backward direction, the output consists of speckle-like noise, demonstrating the model's effectiveness in blocking image formation in the reverse direction.

This joint optimization strategy of the cascaded P-D²NN architecture not only ensures that the structure functions as a cohesive unidirectional image magnification unit but also allows it to be divided into two separate parts, each maintaining its individual unidirectional imaging functionality, as demonstrated in Fig. 8. We conducted individual tests on the unidirectional magnification capabilities of P1 and P2, with the results displayed in Figs. 8b and 8c, respectively. Both diffractive models successfully magnified the input images while blocking the image transmission in the opposite direction, affirming that the smaller models operate effectively as standalone unidirectional image magnifiers. This capability to cascade P-D²NNs demonstrates the potential to achieve larger magnification factors by assembling multiple smaller diffractive models, with significantly less number of diffractive features. For instance, a uniformly-sized standard D²NN would require approximately 97% more diffractive features if its layer size matches the size of P1's last layer for unidirectional image magnification with $M = 3$, and about 1678% more features to match the size of P2's last layer for $M = 9$.

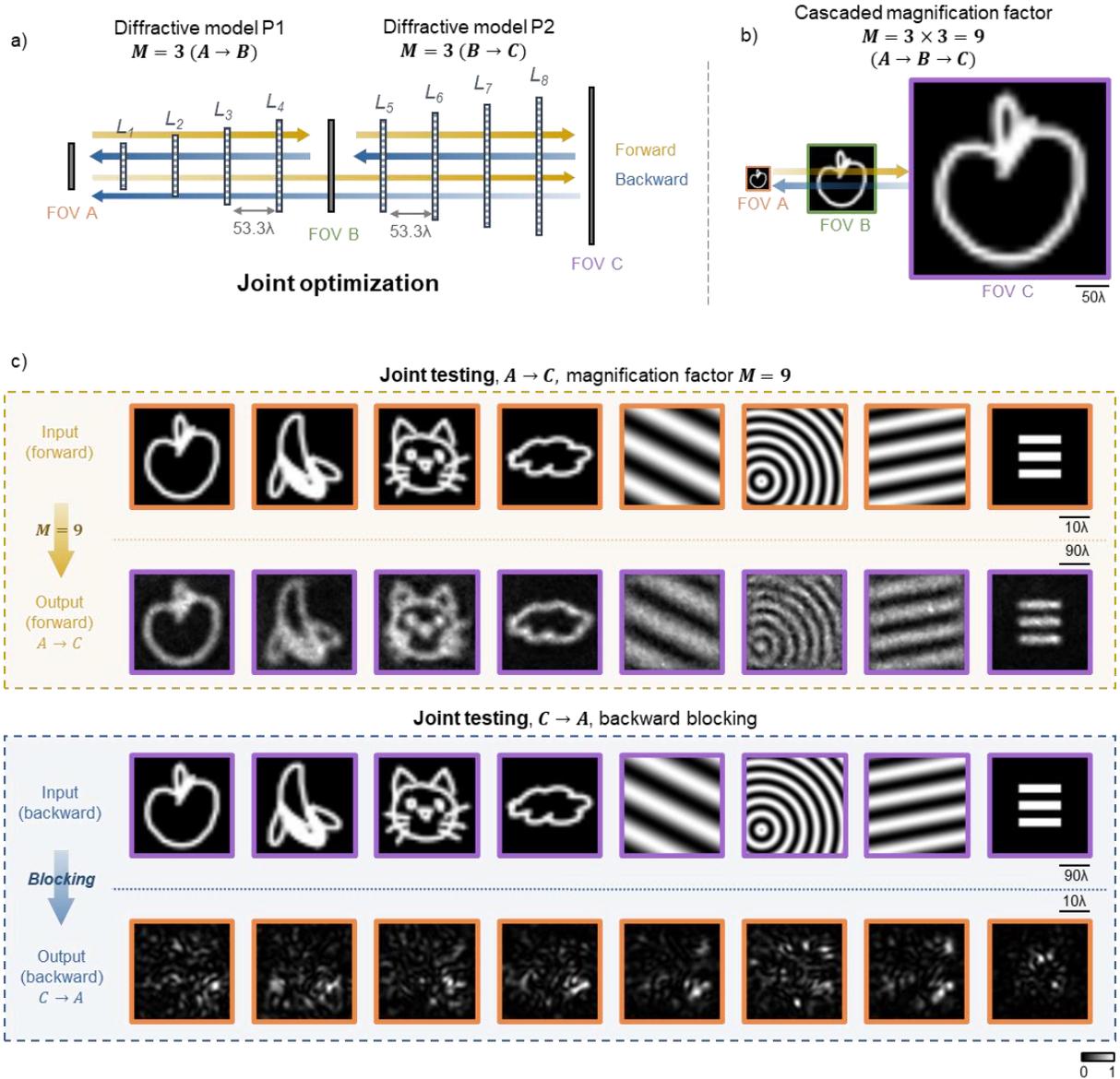

**Figure 7. Cascaded P-D²NN with a joint optimization strategy.** (a) The design concept of the cascaded pyramid diffractive network: two P-D²NN units (P1 and P2) are cascaded, where the output plane of P1 serves as the input plane for P2. The system is optimized using three distinct loss functions—one for each P-D²NN and one for the end-to-end system performance (see the Methods). (b) Cascaded image magnification and field-of-view sizes. The color coding on the boundary represents the size of the FOVs and is consistent for all the cascaded designs. (c) Joint testing of the cascaded P-D²NN architecture with a magnification factor of $M = 3 \times 3 = 9$. In the forward direction, the cascaded network projects a magnified version of the input image, while in the backward direction, only speckle-like noise is observed, blocking the image formation – as desired. All the images are individually normalized.

Note that if we were to optimize the P-D$^2$NN architecture only in an end-to-end manner, without constraints on the individual diffractive components, joint testing of the cascaded network would still demonstrate that the system functions effectively for $M = 9$, as illustrated in Supplementary Fig. S8b. However, when disassembled, neither P1 nor P2 would be able to form a magnified image in the forward direction (see Supplementary Figs. S8c-d). On the other hand, as illustrated in Supplementary Fig. S9, if P1 and P2 are optimized separately and then cascaded without any end-to-end optimization, the assembled structure fails to successfully reconstruct the magnified input images. These results highlight the importance of our joint optimization strategy for the cascaded P-D$^2$NN architecture demonstrated in Fig. 7.

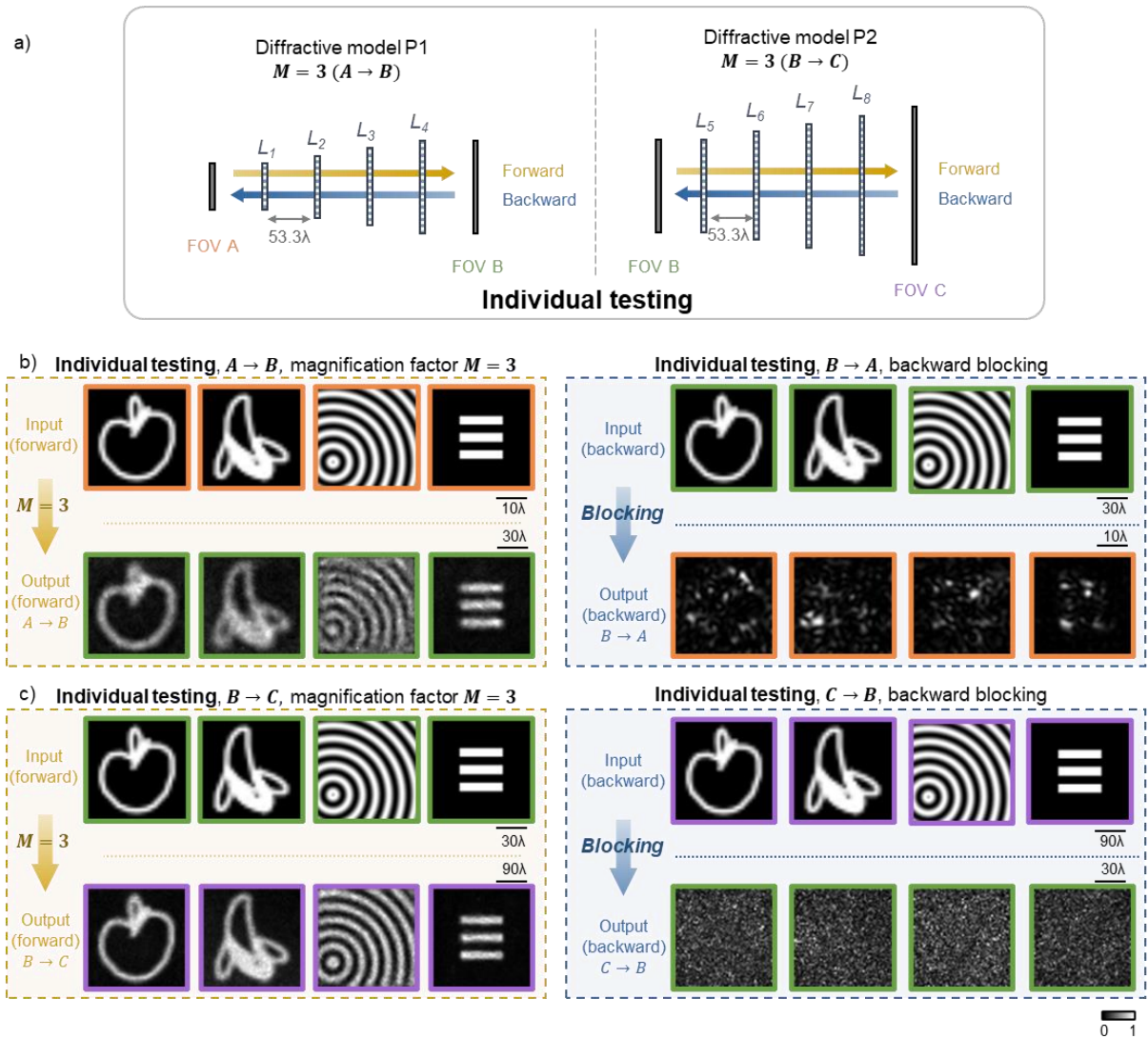

**Figure 8. Individual testing for the cascaded P-D$^2$NN.** (a) The cascaded diffractive structure is split into two individual parts and are separately tested. (b, c) Individual testing results for P1 and P2. All the images are individually normalized.

**Experimental demonstration of a unidirectional magnifier P-D²NN**

We experimentally demonstrated our P-D²NN-based unidirectional image magnifier and demagnifier designs using monochromatic terahertz illumination at $\lambda = 0.75$ mm, as shown in Figs. 9 and 10 (also see the Methods section). For the unidirectional image magnification experimental validation, we constructed a pyramid magnifier consisting of three diffractive layers ($L_1, L_2,$ and $L_3$ in Fig. 9a), where each layer contains $40 \times 40$, $60 \times 60$, and $80 \times 80$ diffractive features, respectively. For the demagnification design, we used the same setup as Fig. 9a but reverted the number of trainable diffractive features on each layer to $80 \times 80$, $60 \times 60$, and $40 \times 40$. Each diffractive feature had a lateral size of ~0.67$\lambda$, selected based on the resolution of our 3D printer. The total length of our experimental setup along the propagation direction is ~26.7$\lambda$ excluding the input and output apertures, and ~53.3$\lambda$ when including them. The pyramid unidirectional magnifier was trained to perform unidirectional image magnification with $M = 2$ in the forward direction, and the unidirectional demagnifier was trained to perform unidirectional image demagnification with $D = 2$ in the forward direction. After the training was completed (see the Methods section for details), the resulting diffractive layers were fabricated using 3D printing and assembled to form the physical unidirectional imager for the THz experimental set-up. The optimized phase modulation maps and the corresponding images of the fabricated layers are shown in Figs. 9b-c and Figs. 10a-b. Additionally, we utilized 3D printing to create customized housings for the diffractive layers, ensuring their correct alignment under experimental conditions. An aluminum coating was also applied to all areas surrounding the diffractive features to block any unwanted light propagation and minimize undesired light coupling.

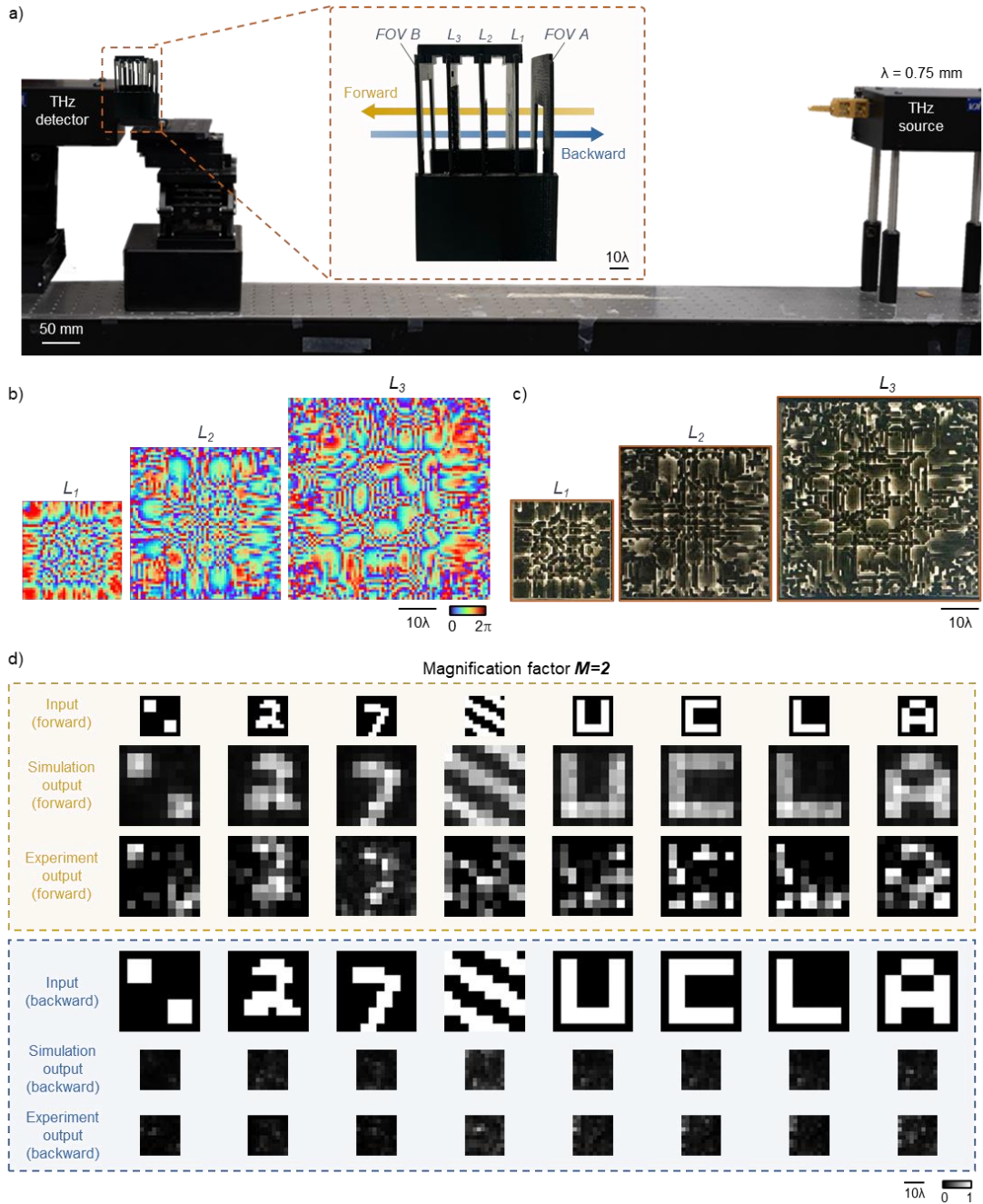

**Figure 9. Experimental demonstration of the unidirectional image magnifier using a pyramid diffractive optical network.** (a) Photographs of the fabricated P-D²NN and the experimental setup with λ = 0.75 mm THz illumination. (b) The converged phase patterns of the diffractive layers. (c) Photographs of the 3D printed diffractive layers with back illumination. (d) Experimental results of the unidirectional magnifier using the fabricated P-D²NN.

In our experiments, we first evaluated the performance of two 3D-printed pyramid unidirectional devices: a magnifier and a demagnifier with $M = 2$ and $D = 2$, respectively. Both devices were tested in the forward and backward directions using several test objects that were not included in the training data. The experimental results are displayed in Fig. 9d for the unidirectional image magnifier, and in Fig. 10c for the unidirectional image demagnifier, alongside their respective numerical testing results. These experimental results confirm that both devices performed as desired. Specifically, the unidirectional magnifier (Fig. 9d) effectively magnified the input images in the forward direction while inhibiting image formation in the backward direction, closely matching our numerical simulations. Similarly, the unidirectional demagnifier (Fig. 10c) successfully reduced the size of the input image in its forward direction and prevented image formation in the backward direction. Our experimental results illustrate a good agreement with the corresponding numerical simulations, demonstrating the proof-of-concept of the 3D-printed P-D²NN designs.

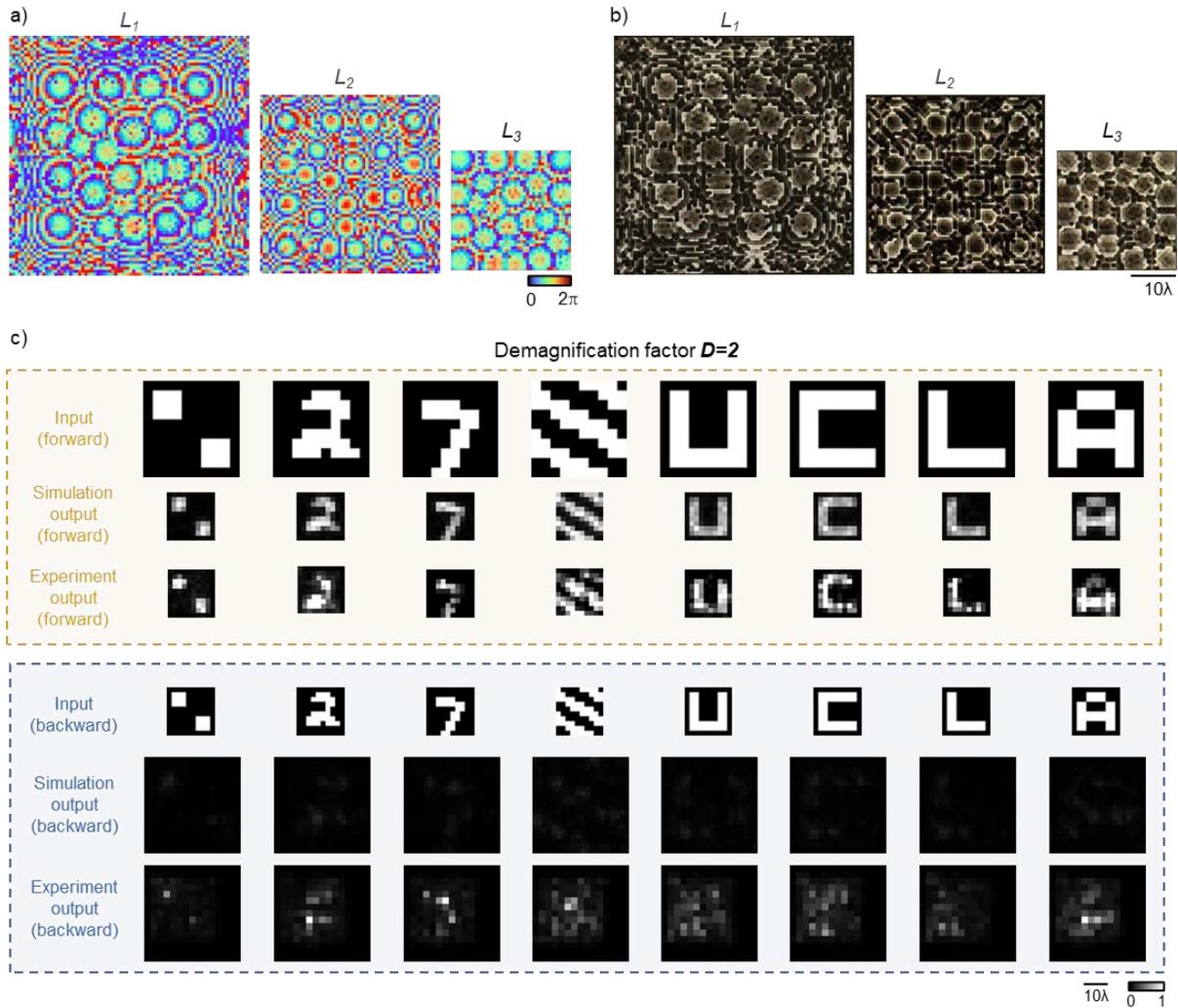

**Figure 10. Experimental demonstration of the unidirectional image demagnifier (D=2) using a pyramid diffractive optical network.** (a) The converged phase patterns of the diffractive layers and (b)

To further explore the capabilities of the P-D²NN, we designed and evaluated a two-layer diffractive model for $M = 3$ (refer to the Methods section for details). Figure 11a illustrates the experimental setup; the optimized phase modulation maps of the resulting diffractive layers, along with the corresponding images of the 3D fabricated structures are shown in Figs. 11b and 11c, respectively. We tested input objects that were not part of the training set, and the experimental results are displayed in Fig. 11d. The P-D²NN system successfully magnified the input images in the forward direction by a factor of $M = 3$, closely matching the numerical simulations, while only noise patterns were observed in the backward direction at the output plane – as desired from a unidirectional imaging system.

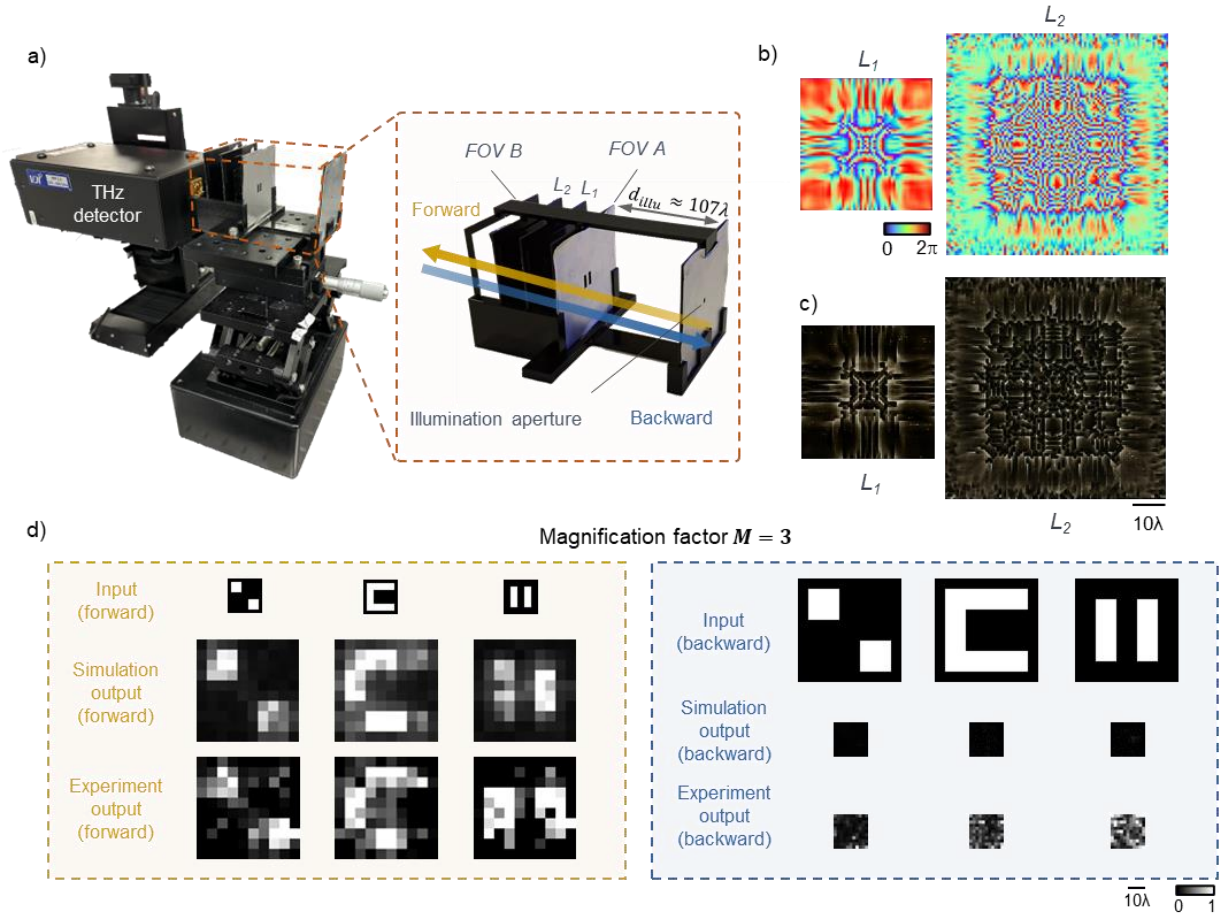

**Figure 11. Experimental demonstration of the unidirectional image magnifier (M=3) using a pyramid diffractive optical network.** (a) The image of the experimental set-up and design. (b) The converged phase patterns of the diffractive layers, and (c) back-illuminated photographs of the 3D printed diffractive layers. (d) Experimental results of the unidirectional magnifier using the fabricated P-D²NN with a magnification factor of M=3.

## Discussion

We presented a pyramid diffractive network architecture where the effective diffraction area scales in alignment with the geometrical scaling operation/task. Compared to conventional uniform-sized $D^2NN$ designs, P-$D^2NN$ learns unidirectional image scaling operations (magnification/demagnification) in a more efficient way by limiting its possible solution space to a confined region that is pre-determined according to the behavior of ray optics. This allows the pyramid diffractive network architecture to converge to a more optimal solution, achieved with fewer diffractive degrees of freedom compared to regular $D^2NN$ designs, where each layer has the same number of diffractive features. In specific tasks, such as unidirectional image magnification, most of the optical energy is transmitted along a defined cone. As the input light diffracts through the P-$D^2NN$ layers, the majority of the energy remains confined within the areas delineated by geometrical optics. Allocating trainable diffractive features within these areas ensures more effective energy utilization. We believe that this physics-inspired approach that integrates task specificity into the structure of the diffractive network layers can foster more efficient visual processors and more optimal task-specific diffractive networks.

As an end-to-end fully differentiable system, P-$D^2NN$ is highly versatile and can be tailored to various desired functionalities through the proper design of loss functions. Similar to a standard imaging system, various forms of aberrations can be taken into account depending on the desired resolution and effective numerical aperture. Another approach to enhance resolution involves incorporating resolution test targets or gratings of various periods into the training dataset, which would further improve the system's imaging performance. Furthermore, P-$D^2NN$ framework can be optimized to generate virtual images at the output aperture, alongside real images. By altering the loss function with respect to the diffracted version of any virtual plane of interest, the image field at the output aperture can be made to appear as if it is diffracting from a desired virtual plane. It is important to note that while other diffractive imaging systems, such as Fresnel zone plates[37], diffractive optical elements[38] and metasurfaces[39] are also optimizable to provide image magnification/demagnification, they lack the unidirectional imaging feature of P-$D^2NN$s, where the image formation is blocked in the reverse direction, distinguishing the P-$D^2NN$ framework from the other image magnification/demagnification systems.

We should note that our P-$D^2NN$ framework is a reciprocal system with asymmetrically structured materials that are linear and isotropic; it does not have time reversal symmetry due to its engineered losses. As an alternative, one can design *nonreciprocal* systems through e.g., the magneto-optic effect[40,41], spatio-temporal modulations[42,43] or nonlinear optical effects[44,45]. However, these approaches have been primarily limited to non-structured, relatively simple beam profiles and are, in general, polarization-sensitive; furthermore, these approaches would be bulky to implement for unidirectional image magnification or demagnification tasks. In contrast, P-D²NN is a polarization-insensitive unidirectional imaging system, with input and output apertures that can consist of millions of pixels once fabricated at a large scale. Therefore, the space bandwidth product (SBP) of the P-D²NN framework can be scaled up to >1 Million through the training and fabrication of larger diffractive layers, potentially offering significant scalability.

For the experimental set-ups, we adopted simpler P-D²NN designs in consideration of potential misalignments during the network assembly, limited signal-to-noise-ratio (SNR) of the THz setup, and other non-ideal experimental conditions. To further understand the impact of some of these factors, we conducted an error analysis to study the effect of phase quantization and fabrication errors on the output image quality. The results of these analyses, summarized in Supplementary Figs. S10 and S11, clearly demonstrate the resilience of the P-D²NN framework to phase quantization and potential fabrication errors (also see the Methods section). These proof-of-concept experiments demonstrated the feasibility of our presented framework, while with more advanced 3D fabrication technologies such as lithography and two-photon polymerization, along with more accurate system alignment and higher SNR sensors, we believe that the gap between the numerical simulations and experimental results can be further improved.

Although the experimental demonstrations of the P-D²NN framework reported in this work were performed under THz illumination, the system is inherently scalable to a broader spectrum of illumination wavelengths, including the infrared (IR) and the visible range. As evident in the spectral response evaluation results reported in Fig. 5, a P-D²NN design, originally trained at a single illumination wavelength, effectively maintains its unidirectional imaging functionality across a significantly extended wavelength range. Therefore, the P-D²NN framework can operate efficiently under broadband illumination. When fabricated in a monolithic fashion using, e.g., two-photon polymerization-based 3D-printers, a P-D²NN design that operates at the visible or IR bands can achieve a very compact footprint, axially spanning <100-200 µm.

Our pyramid design is inspired by not just geometrical optics but also the principles of pruning frequently employed in conventional machine learning[46,47]. The intuition behind pruning also aligns with the idea of Occam's Razor[48] that using a model with redundant degrees of freedom - the regular D²NN in our case - may increase the risk of overfitting, impair optimization efficiency, and ultimately limit the model's generalization capability. Benefitting from this design philosophy, our presented P-D²NN structure can be further tailored for various applications, such as spatial beam shaping and the design of reflective optical processors/components. Moreover, instead of using a fixed architectural design for a given task, the diffractive layer placements and their distributions can be incorporated as trainable parameters and dynamically tuned along with the optimization process. Such an advancement could redefine how diffractive optical networks are constructed, paving the way for task-specific designs that are more efficient and inherently resilient across a spectrum of applications.

## Materials and methods

### Numerical forward model of the diffractive optical network

The pyramid diffractive networks used in this work consist of a series of spatially structured surfaces designed by deep learning, each of which is considered as a thin optical element that modulates only the phase of the transmitted optical field. The transmission coefficient of the trainable diffractive neuron located at $(x, y)$ position of the $k^{th}$ diffractive layer, $t^k$, can be expressed as:

$$t^k(x, y) = \exp\{j\phi^k(x, y)\} \tag{1}$$

where $\phi^k(x, y)$ denotes the phase modulation of the diffractive neuron. Any two consecutive planes are connected to each other by free-space propagation, which is modeled using the angular spectrum approach[10]:

$$u(x, y, z + d) = \mathcal{F}^{-1}\{\mathcal{F}\{u(x, y, z)\} \cdot H(f_x, f_y; d)\} \tag{2}$$

where $u(x, y, z)$ is the original optical field, and $u(x, y, z + d)$ is the resulting field after propagation in free space for a distance of $d$ along the optical axis. $\mathcal{F}$ and $\mathcal{F}^{-1}$ represent the 2D Fourier transform and 2D inverse Fourier transform operations, respectively. $f_x$ and $f_y$ represent the spatial frequencies along the $x$ and $y$ directions, respectively. $H(f_x, f_y; d)$ is the free-space transfer function, which is given by:

$$H(f_x, f_y; d) = \begin{cases} \exp\left\{jkd\sqrt{1 - \left(\dfrac{2\pi f_x}{k}\right)^2 - \left(\dfrac{2\pi f_y}{k}\right)^2}\right\}, & f_x^2 + f_y^2 < \dfrac{1}{\lambda^2} \\ 0, & f_x^2 + f_y^2 \geq \dfrac{1}{\lambda^2} \end{cases} \tag{3}$$

where $\lambda$ is the illumination wavelength, $k = \dfrac{2\pi}{\lambda}$ and $j = \sqrt{-1}$.

By alternatingly applying the operations of free-space propagation (Eq. 2) and diffractive phase modulation (Eq. 1), the resulting complex field at the diffractive network's output can be obtained for a given optical field at the input FOV.

**Training loss functions**

In a general diffractive optical network that performs unidirectional magnification at a factor of $M$, the image magnification is permitted in one specified direction (e.g., forward direction, $A \rightarrow B$), while the image formation is restrained in the reverse direction (e.g., backward direction, $B \rightarrow A$). Consequently, the operations of the diffractive network, in both the image magnification and image blocking directions, can be expressed as,

$$O_{Mag} = \mathbf{D^2NN_{Mag}}(I) \tag{4}$$

$$O_{Blk} = \mathbf{D^2NN_{Blk}}(I_M) \tag{5}$$

where $I$ denotes the input intensity image to be magnified, with $O_{Mag}$ being the output intensity after the diffractive network's modulation in the image magnification direction. Conversely, in the image-blocking direction of the D²NN, $O_{Blk}$ is the resulting image after the network's modulation of the input image $I_M$. $I_M$ is the magnified version of image $I$ with a magnification factor of $M > 1$, which is obtained by resizing the image $I$ by $M$ times using the nearest neighbor interpolation, i.e.,

$$I_M = \mathbf{Resize}(I, M) \tag{6}$$

Note that for a unidirectional imager design both $\mathbf{D^2NN_{Mag}}$ and $\mathbf{D^2NN_{Blk}}$ utilize the same set of diffractive layers. The perspective of the input and output images aligns with the direction of the illumination beam. As the illumination direction switches between the image magnification direction and the image blocking direction, the images flip from left to right.

To optimize a diffractive network-based unidirectional image magnifier, we minimize a set of customized loss functions, defined as,

$$\mathcal{L}\big(I, G_{Mag}, O_{Mag}, O_{Blk}\big) = \mathcal{L}_{Scl}\big(I, G_{Mag}, O_{Mag}\big) + \mathcal{L}_{Blk}\big(O_{Blk}\big) + \mathcal{L}_{Ratio}\big(O_{Mag}, O_{Blk}\big) \tag{7}$$

where $G_{Mag}$ is the ground truth image in the image magnification direction, which is the geometrically magnified version of the input image $I$ with a scaling factor of $M$, i.e.,

$$G_{Mag} = \mathbf{Resize}(I, M) \tag{8}$$

The loss term $\mathcal{L}_{Scl}(\cdot)$ in Eq. 7 is designed to enhance the image magnification (geometrical scaling) fidelity and the energy efficiency in the image magnification direction, which is formulated as,

$$\mathcal{L}_{Scl}(I, G_{Mag}, O_{Mag}) = \mathbf{NMSE}(G_{Mag}, O_{Mag}) + \alpha\left(1 - \mathbf{PCC}(G_{Mag}, O_{Mag})\right) - \beta \exp\left(\boldsymbol{\eta}_{Scl}(I, O_{Mag})\right) \quad (9)$$

where $\alpha$ and $\beta$ are constants that balance the weights of each loss term. $\mathbf{NMSE}(\cdot)$ is the Normalized Mean Square Error, defined as,

$$\mathbf{NMSE}(G_{Mag}, O_{Mag}) = \frac{1}{T}\sum\left(\frac{O_{Mag}}{\max(O_{Mag})} - G_{Mag}\right)^2 \quad (10)$$

where $T$ represents the total number of pixels in each image.

$\mathbf{PCC}(\cdot)$ is the Pearson Correlation Coefficient, defined as,

$$\mathbf{PCC}(G_{Mag}, O_{Mag}) = \frac{\sum(G_{Mag} - \overline{G_{Mag}})(O_{Mag} - \overline{O_{Mag}})}{\sqrt{\sum(G_{Mag} - \overline{G_{Mag}})^2 \sum(O_{Mag} - \overline{O_{Mag}})^2}} \quad (11)$$

where $\overline{G_{Mag}}$ and $\overline{O_{Mag}}$ are the mean values of the intensity images $G_{Mag}$ and $O_{Mag}$, respectively.

$\boldsymbol{\eta}_{Scl}(\cdot)$ is the optical diffraction efficiency along the magnification direction of the diffractive network, which quantifies the ratio of the total energy at the output FOV to the total energy at the input FOV. It is defined as,

$$\boldsymbol{\eta}_{Scl}(I, O_{Mag}) = \frac{\sum O_{Mag}}{\sum I} \quad (12)$$

The loss term $\mathcal{L}_{Blk}(\cdot)$ in Eq. 7 is designed to suppress the intensity/energy of the output image in the image blocking direction, which is formulated as,

$$\mathcal{L}_{Blk}(O_{Blk}) = \gamma \sum_n \mathbf{max_n}(O_{Blk}) \quad (13)$$

which measures the total energy of the top $n$ pixels with the highest intensity values of $O_{Blk}$. $n$ is a hyperparameter that was selected as 50. $\gamma$ is a weighting constant.

The loss term $\mathcal{L}_{Ratio}(\cdot)$ in Eq. 7 is formulated as,

$$\mathcal{L}_{Ratio}(O_{Mag}, O_{Blk}) = \mu \frac{\sum O_{Blk}}{\sum O_{Mag}} \quad (14)$$

which calculates the ratio of total energy at the output FOV in the image blocking direction to that in the image magnification direction, and $\mu$ is a weighting constant. Minimizing $\mathcal{L}_{Ratio}$ enables both the enhancement of the diffraction efficiency along the image magnification direction and the suppression of the diffraction efficiency along the opposite, image blocking direction.

Similarly, in the case of unidirectional image *demagnification*, the diffractive network performs image demagnification in one direction and image blocking in the opposite direction. With a demagnification factor of $D$, the operations of the diffractive network can be expressed as,

$$O_{Demag} = \mathbf{D^2NN_{Demag}}(I) \tag{15}$$

$$O_{Blk} = \mathbf{D^2NN_{Blk}}(I_D) \tag{16}$$

where $I$ is the input intensity image and $O_{Demag}$ is the network's output in the demagnification direction. In the opposite, image-blocking direction (Eq. 16), $I_D$ is the input intensity image and $O_{Blk}$ is the network's output. $I_D$ is the demagnified version of the image $I$ with a demagnification factor of $D > 1$, denoted as:

$$I_D = \mathbf{Resize}(I, 1/D) \tag{17}$$

The loss function used to optimize a unidirectional demagnifier can be written as,

$$\boldsymbol{\mathcal{L}}\big(I, G_{Demag}, O_{Demag}, O_{Blk}\big) = \boldsymbol{\mathcal{L}_{Scl}}\big(I, G_{Demag}, O_{Demag}\big) + \boldsymbol{\mathcal{L}_{Blk}}(O_{Blk}) + \boldsymbol{\mathcal{L}_{Ratio}}\big(O_{Demag}, O_{Blk}\big) \tag{18}$$

where $G_{Demag}$ is the ground truth image in the image demagnification direction, which is the geometrically demagnified version of the input image $I$ with a factor of $D$, i.e.,

$$G_{Demag} = \mathbf{Resize}(I, 1/D) \tag{19}$$

The loss terms $\boldsymbol{\mathcal{L}_{Scl}}(\cdot)$, $\boldsymbol{\mathcal{L}_{Blk}}(\cdot)$, and $\boldsymbol{\mathcal{L}_{Ratio}}(\cdot)$ are the same as defined in Eq. 9, Eq. 13, and Eq. 14.

For the unidirectional magnification network models that are trained under a single illumination wavelength (e.g., in Figs. 2 and 4), the image magnification is designed to be maintained in the forward direction ($A \rightarrow B$) while being suppressed in the backward direction ($B \rightarrow A$). We denote the input, ground truth and output images of the diffractive network in the $A \rightarrow B$ direction as $I_A$, $G_{A \rightarrow B}$, and $O_{A \rightarrow B}$, respectively, and denote the output images of the diffractive network in the $B \rightarrow A$ direction as $O_{B \rightarrow A}$. Based on these definitions, the loss function in Eq. 7 becomes,

$$\boldsymbol{\mathcal{L}}\big(I, G_{Mag}, O_{Mag}, O_{Blk}\big) = \boldsymbol{\mathcal{L}}\big(I = I_A, G_{Mag} = G_{A \rightarrow B}, O_{Mag} = O_{A \rightarrow B}, O_{Blk} = O_{B \rightarrow A}\big) \tag{20}$$

Following the same notation, the loss function for the unidirectional image demagnification network (Eq. 18) trained under a single illumination wavelength (e.g., in Fig. 3) becomes,

$$\mathcal{L}(I, G_{Demag}, O_{Demag}, O_{Blk}) = \mathcal{L}(I = I_A, G_{Demag} = G_{A \to B}, O_{Demag} = O_{A \to B}, O_{Blk} = O_{B \to A}) \quad (21)$$

where $G_{A \to B}$ is the magnified version of $I_A$ in the case of unidirectional image magnification (Eq. 20) and the demagnified version of $I_A$ in the case of unidirectional image demagnification (Eq. 21).

In the wavelength-multiplexed diffractive networks reported in Fig. 6, two opposite operations are performed simultaneously by a single diffractive network operating at two distinct wavelengths, $\lambda_1$ and $\lambda_2$. Specifically, at $\lambda_1$, the diffractive network performs image magnification in $A \to B$ direction and image blocking in $B \to A$ direction. At $\lambda_2$ illumination, however, the diffractive network performs image demagnification in $B \to A$ direction and image blocking in $A \to B$ direction. Therefore, the loss function used to train such a wavelength-multiplexed diffractive network can be expressed as a summation of two wavelength-specific sub-terms,

$$\mathcal{L}(I = I_A, G_{Mag} = G_{A \to B, \lambda_1}, O_{Mag} = O_{A \to B, \lambda_1}, O_{Blk} = O_{B \to A, \lambda_1})$$
$$+ \mathcal{L}(I = I_B, G_{Demag} = G_{B \to A, \lambda_2}, O_{Demag} = O_{B \to A, \lambda_2}, O_{Blk} = O_{A \to B, \lambda_2}) \quad (22)$$

where $I_A$ and $I_B$ are the input images at FOV A and FOV B, respectively. $O_{A \to B, \lambda_1}$ and $O_{B \to A, \lambda_1}$ refer to the output images in $A \to B$ and $B \to A$ directions, respectively, at the illumination wavelength of $\lambda_1$. $O_{A \to B, \lambda_2}$ and $O_{B \to A, \lambda_2}$ refer to the output images in $A \to B$ and $B \to A$ directions, respectively, at the illumination wavelength of $\lambda_2$. $G_{A \to B, \lambda_1}$ is the ground truth image in $A \to B$ direction at $\lambda_1$, which is the magnified version of $I_A$ in this design. $G_{B \to A, \lambda_2}$ is the ground truth image in $B \to A$ direction at $\lambda_2$, which is the demagnified version of $I_B$ in this design.

For the unidirectional image magnification and demagnification P-D$^2$NN models used in experimental testing (see Figs. 9-11), an additional loss term was incorporated to enhance the contrast of the output images in the image magnification direction, i.e.,

$$\mathcal{L}_{exp}(I, G_{Mag}, O_{Mag}, O_{Blk}) = \mathcal{L}(I, G_{Mag}, O_{Mag}, O_{Blk}) + \mathcal{L}_{cnt}(G_{Mag}, O_{Mag}) \quad (23)$$

where $\mathcal{L}(I, G_{Mag}, O_{Mag}, O_{Blk})$ is the same as defined in Eq. 7, and $\mathcal{L}_{cnt}(\cdot)$ is defined as,

$$\mathcal{L}_{cnt}(G_{Mag}, O_{Mag}) = \frac{\sum \left( O_{Mag} \cdot \left( 1 - \hat{G}_{Mag} \right) \right)}{\sum \left( O_{Mag} \cdot \hat{G}_{Mag} \right)} \quad (24)$$

where $\hat{I}_M$ represents a binary mask that identifies the transmissive regions of the input object $I_M$, i.e.,

$$\hat{G}_{Mag}(x, y) = \begin{cases} 1, & G_{Mag}(x, y) > 0.5 \\ 0, & otherwise \end{cases} \quad (25)$$

**Quantification metrics used for performance testing**

To quantify the performance of our unidirectional image magnifier/demagnifier designs, the PCC values between the output and ground truth images (in both the forward and backward directions), the diffraction efficiency (in both forward and backward directions), and the energy ratio of the output images in the forward direction to the backward direction were selected as quantitative figures of merits. Specifically, the PCC value in the forward direction ($A \rightarrow B$) or the backward direction ($B \rightarrow A$) can be calculated as,

$$\text{Forward PCC} = \textbf{PCC}_{A \rightarrow B}(G_{A \rightarrow B}, O_{A \rightarrow B}) \tag{26}$$

$$\text{Backward PCC} = \textbf{PCC}_{B \rightarrow A}(G_{B \rightarrow A}, O_{B \rightarrow A}) \tag{27}$$

where $\textbf{PCC}(\cdot)$ is as defined in Eq. 11. $G_{A \rightarrow B}$ and $G_{B \rightarrow A}$ are the ground truth images in $A \rightarrow B$ and $B \rightarrow A$ directions, respectively. In the case of unidirectional magnification (e.g., in Fig. 2), $G_{A \rightarrow B}$ is the magnified version of the input image $I_A$. In the case of unidirectional demagnification (e.g., in Fig. 3), $G_{A \rightarrow B}$ is the demagnified version of the input image $I_A$. $G_{B \rightarrow A}$ is the resized (magnified/demagnified) version of $I_B$. $O_{A \rightarrow B}$ and $O_{B \rightarrow A}$ refers to the output images in the $A \rightarrow B$ and $B \rightarrow A$ directions, respectively.

Similarly, the diffraction efficiency in the forward ($A \rightarrow B$) or backward directions ($B \rightarrow A$) can be calculated as,

$$\text{Forward diffraction efficiency} = \boldsymbol{\eta}_{A \rightarrow B}(I_A, O_{A \rightarrow B}) = \frac{\sum O_{A \rightarrow B}}{\sum I_A} \tag{28}$$

$$\text{Backward diffraction efficiency} = \boldsymbol{\eta}_{B \rightarrow A}(I_B, O_{B \rightarrow A}) = \frac{\sum O_{B \rightarrow A}}{\sum I_B} \tag{29}$$

Finally, the forward-backward energy ratio can be calculated as,

$$\text{Forward} - \text{backward energy ratio} = \frac{\sum O_{A \rightarrow B}}{\sum O_{B \rightarrow A}} \tag{30}$$

The FWHM values are calculated based on the gradient of the line-spread functions as:

$$\text{FWHM} = |x_2 - x_1| \tag{31}$$

where $x_1$ and $x_2$ are the solutions of

$$f_{PSF}(x) = \frac{\max\{f_{PSF}\}}{2} \tag{32}$$

Here $f_{PSF}$ is calculated as the gradient of the line-spread function. The line-spread functions are calculated by averaging over 11 cross-sections evenly spaced within the FOV. The final FWHM reported is averaged over the 9 images with different angles (see Supplementary Fig. S5).

## Digital implementation and training details

The diffractive network models used in our numerical simulations have a diffractive feature/neuron size of ~$0.53\lambda$, where $\lambda = 0.75$ mm. The pyramid network for unidirectional image magnification, as reported in Fig. 2, contains five diffractive layers with sequentially increasing numbers of trainable diffractive features on each layer. From the first layer $L_1$ through the fifth layer $L_5$, the diffractive layers progressively increased, with 90×90, 140×140, 180×180, 220×220, and 270×270 diffractive neurons at each layer respectively, leading to a total number of trainable neurons of $N = N_b = 181,400$. The magnification factor in the forward direction was selected as $M = 3$, with an input FOV comprising 90×90 pixels, and the output FOV having 270×270 pixels. The axial distance between any two consecutive planes was set as 40mm (i.e., $53.3\lambda$). The weights of the loss terms used for training were chosen as: $\alpha = 8$, $\gamma = 1$, and $\mu = 2$, with $\beta$ varied across [0.5, 0.8, 1.0, 1.5, 2.0, 4.0] to generate the results reported in Fig. 2d and Supplementary Fig. S1. For Supplementary Fig. S2, the number of trainable diffractive features for models with different $K$ was $[90^2, 150^2, 210^2, 270^2]$ for $K = 4$, $[90^2, 180^2, 270^2]$ for $K = 3$ and $[150^2, 270^2]$ for $K = 2$, while all other parameters remained the same. For the grating and slanted edge testing (Supplementary Figs. S3-5), we used a larger model with five diffractive layers consisting of $[180^2, 210^2, 240^2, 270^2, 300^2]$ diffractive features, with all the other parameters kept the same as the model reported in Fig. 2.

The unidirectional image demagnification pyramid network reported in Fig. 3 adopts a symmetric geometric arrangement with respect to its magnification counterpart (Fig. 2), in which the five diffractive layers have progressively decreasing numbers of trainable neurons as 270×270, 220×220, 180×180, 140×140, and 90×90, respectively. The axial distance between any two consecutive planes was set as 40mm. The demagnification factor in the forward direction was selected as $D = 3$, with an input FOV comprising 270×270 pixels, and the output FOV having 90×90 pixels. The weights of the loss terms used for training were chosen as: $\alpha = 8$, $\gamma = 1$, and $\mu = 2$, with $\beta$ varied across [0.5, 1.0, 1.5, 2.0, 3.0, 4.0] to generate the results reported in Fig. 3d and Supplementary Fig. S6.

The 5-layer regular diffractive network reported in Fig. 4b is designed to achieve unidirectional image magnification at a factor of $M = 3$. The input and output FOVs have 90×90 and 270×270 pixels, respectively. Each of the diffractive layers has 270x270 trainable neurons, summing up to $N = 2N_b$ trainable neurons across the structure. The axial separation between any two consecutive planes was also set as 40 mm (~53.3$\lambda$). The weights in the training loss functions were selected as: $\alpha = 8$, $\beta = 1$, $\gamma = 1$, and $\mu = 2$ to be compared with its pyramid counterpart trained with the same set of weight parameters.

The wavelength-multiplexed diffractive network reported in Fig. 6 retains the same geometric architecture as in Fig. 2. The two training wavelengths were selected as $\lambda_1 = 0.75$ mm and $\lambda_2 = 0.8$ mm. The weights in the training loss functions were also selected as: $\alpha = 8$, $\gamma = 1$, and $\mu = 2$, with $\beta$ varied across [1.0, 2.5, 3.0, 4.0, 5.0] to generate the results reported in Fig. 6c.

For the THz experimental verification, the pyramid diffractive network for unidirectional image magnification has a diffractive feature size of 0.5 mm (~0.67$\lambda$). The sampling period of the optical field was chosen as 0.25 mm (~0.33$\lambda$) to ensure precise modeling. The diffractive network consists of three diffractive layers with 40×40, 60×60, and 80×80 diffractive neurons on each layer. The magnification factor in the forward direction was selected as $M = 2$, with the input and output FOVs having the physical sizes of 15 mm × 15 mm and 30 mm × 30 mm, respectively. The input and output FOVs are sampled into arrays of 10×10 pixels, with an individual pixel having a size of 1.5 mm and 3 mm (2$\lambda$ and 4$\lambda$), respectively. The demagnification model utilizes a similar setup as in the magnification model but with the size of the diffractive layers reversed in order. The sizes of the input and output FOVs are also switched accordingly. For the $M = 3$ experimental design, we trained a two-layer diffractive design employing $\beta = 4$ to enhance the system's output energy efficiency. The two layers comprised 60×60 and 100×100 diffractive features, separated by a distance of ~26.7$\lambda$ (20 mm) which is also the distance from the second layer to the sensor plane and from the object plane to the first layer. A square aperture of 3×3 mm is placed ~107$\lambda$ (80 mm) away from the object plane and is used for both forward and backward illumination.

All the diffractive optical network models reported in the paper were trained with the QuickDraw dataset supplemented by a custom-created dataset comprising grating/fringe-like patterns with various linewidths[17,29]. The training data contains 200,000 images with 120,000 from the QuickDraw dataset and 80,000 from our customized dataset. The validation data contains 50,000 images with 30,000 from the QuickDraw dataset and 20,000 from our customized image dataset. The blind testing data contains 1600 images with 1500 from the QuickDraw dataset and 100 from our customized image dataset, without any overlap with the training or validation datasets. Each image was normalized to the range [0, 1], followed by a set of random image transformations (for data augmentation), including image rotation randomly selected from a range between $-10°$ and $+10°$, scaling with a factor sampled within [0.9, 1.1], and a lateral shift in each direction, with values randomly drawn from $[-\lambda, +\lambda]$.

All the diffractive models in this study were trained and tested using PyTorch v1.13 with a GeForce RTX 3090 graphical processing unit (GPU, Nvidia Inc.). All the models were trained using the Adam optimizer[49] for 20 epochs with a learning rate of 0.03. The diffractive models designed under a single illumination wavelength (e.g., Figs. 2-4) were trained with a batch size of 100. The training typically takes ~5 hours for 20 epochs. The diffractive model designed for wavelength-multiplexed operation (e.g., Fig. 6) was trained with a batch size of 50. The training takes ~9 hours for 20 epochs. The diffractive model for experimental demonstration (e.g., Fig. 9) was trained with a batch size of 200. The training takes ~0.5 hours for 20 epochs.

For the cascaded P-D$^2$NN designs, the input, intermediate and output FOVs (i.e., FOVs A, B, and C) have $60×60$, $180×180$, and $540×540$ pixels, respectively. Each individual P-D$^2$NN (P1 and P2) has four diffractive layers, spaced by ~53.3$\lambda$. The distances from the output plane of P1 to the intermediate plane (FOV B) and from there to the first layer of P2 are also maintained at ~53.3$\lambda$. The number of diffractive features for each layer is sequentially set to $[60^2, 100^2, 140^2, 180^2]$ for P1 and $[180^2, 300^2, 420^2, 540^2]$ for P2.

The joint optimization loss function is given by:

$$\mathcal{L}_{joint} = w_{P1}\mathcal{L}_{P1,A\rightarrow B} + w_{P2}\mathcal{L}_{P2,B\rightarrow C} + w_{cascade}\mathcal{L}_{cascade,A\rightarrow C} \tag{33}$$

where, for the joint optimization case shown in Fig. 7, we used $w_{P1} = w_{P2} = w_{cascade} = 1$ while in the end-to-end optimization case (reported in Supplementary Fig. S8), we used $w_{P1} = w_{P2} = 0$ and $w_{cascade} = 1$. For the individual optimization case (reported in Supplementary Fig. S9), we used $w_{P1} = w_{P2} = 1$ and $w_{cascade} = 0$.

The loss term for each individual part (or the entire diffractive structure) in Eq. 31 contains the same components as outlined in Equation 7. For instance, the loss function for the end-to-end optimization from FOV A to C is given by:

$$\boldsymbol{\mathcal{L}_{cascade,A\rightarrow C} = \mathcal{L}_{Scl}(I, G_{Mag}, O_{Mag}) + \mathcal{L}_{Blk}(O_{Blk}) + \mathcal{L}_{Ratio}(O_{Mag}, O_{Blk})} \tag{34}$$

where $I$ is the input at FOV A and $G_{Mag} = \textbf{Resize}(I, 9)$ and $O_{Mag}$ is captured at FOV C and $O_{Blk}$ is captured at FOV A with $G_{Mag}$ being the input at FOV C.

Unless otherwise stated, the hyperparameters for training remain the same as the diffractive model reported in Fig. 2. All the models were trained and evaluated on a high-performance computing cluster equipped with 8× Nvidia A100 GPUs, each featuring 80 GB of VRAM, with a batch size of 96. Each model undergoes training for 30 epochs, requiring ~24 hours to converge.

**Error analysis simulations**

To simulate the impact of phase quantization error at each diffractive feature, we denoted $\phi_{bit}$ as the phase bit depth, covering $2^{\phi_{bit}}$ phase values evenly spaced in $[0, 2\pi)$. We blindly tested an optimized diffractive model ($K = 5$, $\beta = 1$, trained using a single-precision floating format) using limited $\phi_{bit}$ values of 4, 3 and 2 by rounding the phase value of each ideal/designed diffractive feature to the nearest available value; the results of this analysis are reported in Supplementary Fig. S10.

To model the impact of potential fabrication errors, we introduced the fabrication error strength $\tau_{fab}$ where the final fabricated phase map can be written as $\phi_{fab}(x, y) = \phi_{sim}(x, y) \times (\epsilon(x, y) \cdot \tau_{fab} + 1)$, where $\phi_{sim}(x, y)$ is the simulated/designed phase map and $x, y$ are the spatial coordinates. The random variable $\epsilon(x, y) \sim \mathcal{N}(0,1)$ follows a normal distribution. We tested the same optimized model with $\tau_{fab}$ values of 0.01, 0.05, 0.1 and 0.2, indicating progressively increased fabrication errors; the results of this analysis are reported in Supplementary Fig. S11.

**Experimental demonstration under THz radiation**

Figure 9a and Supplementary Fig. S12 illustrate the schematic diagram of the experimental set-up. The incident terahertz wave was generated by a modular amplifier (Virginia Diode Inc. WR9.0M SGX)/multiplier chain (Virginia Diode Inc. WR4.3x2 WR2.2x2) (AMC) with a compatible diagonal horn antenna (Virginia Diode Inc. WR2.2). A 10 dBm RF input signal at 11.1111 GHz ($f_{RF1}$) from the synthesizer (hp 8340B) was multiplied 36 times by the AMC to generate the output continuous-wave (CW) radiation at 0.4 THz. The AMC was modulated with a 1kHz square wave for lock-in detection. The object plane of the 3D-printed diffractive network was placed ~75 cm away from the exit aperture of the horn antenna. The distance is far enough to approximate the incident wave as a plane wave. The output plane of the diffractive network was 2D scanned using a Mixer (Virginia Diode Inc. WRI 2.2) placed on an XY positioning stage built by vertically combining two linear motorized stages (Thorlabs NRT100). For $M = 2$ experiments, we used a 0.75 mm step size for a FOV of 30 mm×30 mm, and for $D = 2$ experiments, we used a step size of 0.5 mm for a FOV of 15 mm×15 mm; for the $M = 3$ experiments, we used a step size of 1 mm for a FOV of 45 mm×45 mm.

A 10 dBm RF signal at 11.0833 GHz ($f_{RF2}$) was sent to the detector as a local oscillator to down-convert the signal to 1 GHz for further measurement. The down-converted signal was amplified by a low-noise amplifier (Mini-Circuits ZRL-1150-LN+) and filtered by a 1 GHz (+/-10 MHz) bandpass filter (KL Electronics 3C40-1000/T10-O/O). The signal was first measured by a low-noise power detector (Mini-Circuits ZX47-60) and read by a lock-in amplifier (Stanford Research SR830) with the 1 kHz square wave as the reference signal. The raw data were calibrated into a linear scale. Digital binning operations were applied to the calibrated data to match the object feature size used in the numerical simulations.

All the layers and holders were 3D-printed with Object30 V5 Pro (Stratasys) using Vero Black Plus material. Note that this material is non-conductive, and the THz wave reflections from the inner walls of the holder are negligible. A photograph of the 3D-printed holder is shown in Supplementary Fig. S13.

**Acknowledgments:**

The Ozcan Research Group at UCLA acknowledges the support of ONR (Grant # N00014-22-1-2016). The Jarrahi Research Group at UCLA acknowledges the support of NSF (Grant # 2141223).

**Conflict of interest:**

The authors declare that they have no competing interests.

**Author contributions:**

A.O. conceived the research and initiated the project. B.B., X.Y., J. L., and D.M. developed numerical simulation codes. B.B. and X.Y. performed numerical simulations. B.B., X.Y., and T.G. performed the fabrication and experimental testing of the diffractive network. All the authors participated in the analysis and discussion of the results. B.B., X.Y., and A.O. prepared the manuscript and all authors contributed to the manuscript. A.O. supervised the project.

**Data and material availability:**

All the data and methods that support this work are present in the main text and the Supplementary Information. The deep learning models in this work employ standard libraries and scripts that are publicly available in PyTorch.